\documentclass[12pt, a4paper]{article}
\usepackage[latin1]{inputenc}
\usepackage{amsmath}
\usepackage{multicol}
\usepackage{amsfonts}
\usepackage{amssymb}
\usepackage{graphicx}
\usepackage{hyperref}
\usepackage{float}
\usepackage{booktabs}
\usepackage{biblatex}
\usepackage{caption}
\usepackage{diagbox}
\usepackage{mathtools,xparse}

\usepackage[T1]{fontenc}
\usepackage{color}
\usepackage{bm}
\usepackage{enumitem}
\usepackage{booktabs}
\usepackage{algorithm}
\usepackage{algpseudocode}

\setlist[enumerate,1]{label=\textbf{\arabic*.}}
\setlist[enumerate,2]{label=\textbf{\alph*)}}

\title{Theoretical and Empirical Validation of Heston Model}
\author{Zheng Cao \and Xinhao Lin \thanks{Both authors contributed equally to this compilation.}}
\date{Supervisor: Professor Roza Galeeva \\
Johns Hopkins University
}

\begin{document}

\maketitle

\begin{abstract}
    This study focuses on the application of the Heston model to option pricing, employing both theoretical derivations and empirical validations. The Heston model, known for its ability to incorporate stochastic volatility, is derived and analyzed to evaluate its effectiveness in pricing options. For practical application, we utilize Monte Carlo simulations alongside market data from the Crude Oil WTI  market to test the model's accuracy. Machine-learning based optimization methods are also applied for the estimation of the five Heston parameters. By calibrating the model with real-world data, we assess its robustness and relevance in current financial markets, aiming to bridge the gap between theoretical finance models and their practical implementations. 
\end{abstract}


\newpage
\section{Introduction From BS to Heston}

Suggested by Heston in \cite{heston1993closed},  while the Black-Scholes formula, introduced in \cite{black1973pricing}, generally predicts stock option prices effectively, it exhibits biases and performs poorly on derivatives such as foreign currency options. In addition, the Black-Scholes model assumes a constant volatility of the underlying asset, which is disagreed by the market phenomenon of volatility smile curve. To address these limitations, modifications involving stochastic volatility were introduced by Scott \cite{scott1987option}, Hull and White \cite{hull1987pricing}, and Wiggins \cite{wiggins1987option} in 1987. However, these revised models lack closed-form solutions and require complex numerical methods. Furthermore, approaches like those by Jarrow and Eisenberg \cite{eisenberg1994option} and Stein and Stein \cite{stein1991stock} in 1991, which average Black-Scholes values across different volatility paths, fail to account for important skewness effects due to their assumption of uncorrelated volatility and spot returns.

Therefore, Heston suggested a stochastic volatility model, where the volatility follows a mean-reversion stochastic process.  It provides a closed-form solution for the price of a European call option when the spot asset is correlated with volatility, and it adapts the model to incorporate stochastic interest rates.

This report aims to validate the Heston model. In Section \ref{sec:deriv}, we revisited the derivation of Heston model provided in \cite{heston1993closed} and \cite{gatheral2011volatility}. In Section \ref{sec:simu}, we used both crude Monte Carlo method and Monte Carlo method basing on mixing theorem to verify the pricing of Heston model. In Section \ref{sec:greeks_iv}, we derived the five Greeks of Heston model basing on mixing theorem, and we visualized the implied volatility (IV) curves of Heston and analyzed its relation with Heston parameters. In Section \ref{sec:calib}, we calibrate Heston model to WTI future option data from the market regarding some or all of the five Heston parameters as variables. Conclusions about the effectiveness of Heston model and possible areas of future work will be made in Section \ref{sec:conclusion}.

\newpage
\section{Derivation of Heston Model}
\label{sec:deriv}

The Heston model, proposed by \cite{heston1993closed}, describes the evolution of underlying asset price and its variance by
\begin{equation}\label{eqn:hst_model}
    \begin{aligned}\mathrm{d}S(t)&=\mu S(t)\mathrm{d}t+\sqrt{v(t)}S(t)\mathrm{d}W_1(t)\\
    \mathrm{d}v(t)&=-\lambda(v(t)-\bar{v})\mathrm{d}t+\eta\sqrt{v(t)}\mathrm{d}W_2(t)\end{aligned}
\end{equation}
with initial conditions $S(0)=S_0$ and $v(0)=v_0$, where $S(t)$ and $v(t)$ are the underlying price and variance at time $t$, $\mu$ is the physical return of the underlying, $\lambda$ is the mean-reversion speed, $\eta$ is the volatility of volatility, and $W_1(t)$ and $W_2(t)$ are two Brownian motion with correlation $\rho$, i.e., $\mathrm{d}W_1(t)\mathrm{d}W_2(t)=\rho\mathrm{d}t$. In this section, we show the derivation of pricing formula for European options basing on Equation \eqref{eqn:hst_model}.

\subsection{Pricing European Call Options}

Let $c_{\mathrm{H}}(T-t)$ denote the call option value at time $t$ under Heston model. According to the general valuation equation in Chapter 1 of \cite{gatheral2014arbitrage}, we have
\begin{equation}\label{eqn:general_pde}
    \frac{\partial c_{\mathrm{H}}}{\partial t}+\frac{1}{2}vS^2\frac{\partial^2c_{\mathrm{H}}}{\partial S^2}+\rho\eta vS\frac{\partial^2c_{\mathrm{H}}}{\partial v\partial S}+\frac{1}{2}\eta^2v\frac{\partial^2c_{\mathrm{H}}}{\partial v^2}+rS\frac{\partial c_{\mathrm{H}}}{\partial S}-rc_{\mathrm{H}}=\lambda(v-\bar{v})\frac{\partial c_{\mathrm{H}}}{\partial v}
\end{equation}
where $r$ is the risk-free rate. Assume that the option expires at time $T$. By letting $\tau:=T-t$ and $x(t):=\ln{(S(t)e^{r\tau}/K)}$, we can convert Equation \eqref{eqn:general_pde} to
\begin{equation}\label{eqn:hst_pde}
    -\frac{\partial c_{\mathrm{H}}}{\partial\tau}+\frac{1}{2}v\frac{\partial^2c_{\mathrm{H}}}{\partial x^2}-\frac{1}{2}v\frac{\partial c_{\mathrm{H}}}{\partial x}+\rho\eta v\frac{\partial^2c_{\mathrm{H}}}{\partial v\partial x}+\frac{1}{2}\eta^2v\frac{\partial^2c_{\mathrm{H}}}{\partial v^2}-rc_{\mathrm{H}}-\lambda(v-\bar{v})\frac{\partial c_{\mathrm{H}}}{\partial v}=0.
\end{equation}
According to \cite{duffie2000transform}, the solution to Equation \eqref{eqn:hst_pde} has the form
\begin{equation}\label{eqn:hst_form}
    c_{\mathrm{H}}(\tau)=S_0P_1(x,v,\tau)-Ke^{-r\tau}P_0(x,v,\tau).
\end{equation}
Substituting Equation \eqref{eqn:hst_form} into Equation \eqref{eqn:hst_pde}, we get the PDEs for $P_0$ and $P_1$ as
\begin{equation}
    -\frac{\partial P_j}{\partial\tau}+\frac{1}{2}v\frac{\partial^2P_j}{\partial x^2}-\left(\frac{1}{2}-j\right)v\frac{\partial P_j}{\partial x}+\frac{1}{2}\eta^2v\frac{\partial^2P_j}{\partial v^2}+\rho\eta v\frac{\partial^2P_j}{\partial x\partial v}+(\lambda\bar{v}-b_jv)\frac{\partial P_j}{\partial v}=0
\end{equation}
for $j=0, 1$, where $b_j=\lambda-j\rho\eta$, subject to the terminal condition
\begin{equation}\label{eqn:terminal_P}
    \lim_{\tau\rightarrow0}P_j(x,v,\tau)=\begin{cases}
        1, \quad\text{if } x>0\\ 0, \quad\text{if } x\leq0
    \end{cases}:=\theta(x).
\end{equation}
Define the Fourier transform of $P_j$ as
\begin{equation}
    \tilde{P}(u,v,\tau)=\int_{-\infty}^{\infty}e^{-iux}P(x,v,\tau)\mathrm{d}x
\end{equation}
and we have $\tilde{P}(u,v,0)=\frac{1}{iu}$. The inverse transform is
\begin{equation}
    P(x,v,\tau)=\int_{-\infty}^{\infty}\frac{1}{2\pi}e^{iux}\tilde{P}(u,v,\tau)\mathrm{d}u.
\end{equation}
Substituting into Equation \eqref{eqn:hst_pde}, we get
\begin{equation}\label{eqn:pde_tilde_P}
    \begin{aligned}
        &-\frac{\partial\tilde{P}_j}{\partial\tau}-\frac{1}{2}u^2v\tilde{P}_j-\left(\frac{1}{2}-j\right)iuv\tilde{P}_j+\frac{1}{2}\eta^2v\frac{\partial^2\tilde{P}_j}{\partial v^2}+\rho\eta iuv\frac{\partial\tilde{P}_j}{\partial v}+(\lambda\bar{v}-b_jv)\frac{\partial\tilde{P}_j}{\partial v}\\
        =&v\left(\alpha\tilde{P}_j-\beta\frac{\partial\tilde{P}_j}{\partial v}+\gamma\frac{\partial^2\tilde{P}_j}{\partial v^2}\right)+\lambda\bar{v}\frac{\partial\tilde{P}_j}{\partial v}-\frac{\partial\tilde{P}_j}{\partial\tau}=0
    \end{aligned}
\end{equation}
where $\alpha=-\frac{u^2}{2}-\frac{iu}{2}+iju$, $\beta=\lambda-\rho\eta j-\rho\eta iu$, and $\gamma=\frac{\eta^2}{2}$. Now we search for the solution to Equation \eqref{eqn:pde_tilde_P} with the form
\begin{equation}
    \tilde{P}_j(u,v,t)=\exp\left\{C(u,\tau)\bar{v}+D(u,\tau)v\right\}\tilde{P}_j(u,v,0)=\frac{1}{iu}\exp\left\{C(u,\tau)\bar{v}+D(u,\tau)v\right\}.
\end{equation}
Substituting into Equation \eqref{eqn:pde_tilde_P}, we get
\begin{equation}
    \begin{aligned}
        \frac{\partial C}{\partial\tau}&=\lambda D\\
        \frac{\partial D}{\partial\tau}&=\alpha-\beta D+\gamma D^2
    \end{aligned}.
\end{equation}
Integrating with $C(u,0)=D(u,0)=0$, we get
\begin{equation}
    \begin{aligned}
        D(u,\tau)&=r_-\frac{1-e^{-d\tau}}{1-ge^{-d\tau}}\\
        C(u,\tau)&=\lambda\left[r_-\tau-\frac{2}{\eta^2}\ln\left(\frac{1-ge^{-d\tau}}{1-g}\right)\right]
    \end{aligned}
\end{equation}
where $r_{\pm}=\frac{\beta\pm\sqrt{\beta^2-4\alpha\gamma}}{2\gamma}=:\frac{\beta\pm d}{\eta^2}$ and $g:=\frac{r_-}{r_+}$. Then, taking the inverse transform, we get
\begin{equation}\label{eqn:P_j_call}
    P_j(x,v,\tau)=\frac{1}{2}+\frac{1}{\pi}\int_0^{\infty}Re\left\{\frac{\exp\{C_j(u,\tau)\bar{v}+D_j(u,\tau)v+iux\}}{iu}\right\}\mathrm{d}u.
\end{equation}

\subsection{Pricing European Put Options}

In the pricing of put options, the terminal condition given in Equation \eqref{eqn:terminal_P} becomes
\begin{equation}\label{eqn:terminal_P_put}
    \lim_{\tau\rightarrow0}P_j(x,v,\tau)=\begin{cases}
        0, \quad\text{if } x>0\\ -1, \quad\text{if } x\leq0
    \end{cases}.
\end{equation}
Hence, through the same derivation process above, the solution for $P_j$ is
\begin{equation}\label{eqn:P_j_put}
    P_j(x,v,\tau)=-\frac{1}{2}+\frac{1}{\pi}\int_0^{\infty}Re\left\{\frac{\exp\{C_j(u,\tau)\bar{v}+D_j(u,\tau)v+iux\}}{iu}\right\}\mathrm{d}u.
\end{equation}
Given the call option price (combining Equation \eqref{eqn:hst_form} and \eqref{eqn:P_j_call}) and put option price (combining Equation \eqref{eqn:hst_form} and \eqref{eqn:P_j_put}), one can easily check the put-call parity given by
\begin{equation}\label{eqn:parity}
    c_{\mathrm{H}}-p_{\mathrm{H}}=S_0-Ke^{-r\tau}
\end{equation}
where $c_{\mathrm{H}}$ and $p_{\mathrm{H}}$ are the Heston call and put option prices under the same parameters setting, correspondingly.

\subsection{Extreme Case: Deterministic Variance}
\label{sec:eta0}

If $\eta=0$, meaning that the variance $v(t)$ in Equation \eqref{eqn:hst_model} is deterministic, then we can solve for $v(t)$ via the ODE
\begin{equation}
    \frac{dv(t)}{dt}=-\lambda(v(t)-\bar{v})
\end{equation}
with initial condition $v(0)=v_0$. The solution is
\begin{equation}
    v(t)=\bar{v}+(v_0-\bar{v})e^{-\lambda t}.
\end{equation}
Then, according to the generalized Black-Scholes model, the call option price is
\begin{equation}
    c_{\mathrm{H}}=c_{\mathrm{BS}}(S,K,r,T,\sigma^*)
\end{equation}
where $\sigma^*$ is the equivalent volatility calculated by
\begin{equation}\label{eqn:sgm_eta0}
    \sigma^*=\sqrt{\frac{\int_0^Tv(t)dt}{T}}=\sqrt{\bar{v}+\frac{1-e^{-\lambda T}}{\lambda T}(v_0-\bar{v})}
\end{equation}
and $c_{\mathrm{BS}}(\cdot)$ is the Black-Scholes formula for call option calculated by
\begin{equation}\label{eqn:bs_call}
c_{\mathrm{BS}}(S, K, r, T, \sigma)=S\Phi(d_+)-Ke^{-rT}\Phi(d_-)
\end{equation}
where $d_{\pm}=\frac{\ln{\frac{S}{K}}+(r\pm\frac{\sigma^2}{2})t}{\sigma\sqrt{t}}$ and $\Phi(x)=\frac{1}{\sqrt{2\pi}}\int_{-\infty}^xe^{-\frac{x^2}{2}}dx$.

\newpage
\section{Simulation}
\label{sec:simu}
In this section, we discuss our implementation of two simulation methods: crude Monte Carlo (MC) and mixing MC. We verify the pricing of Heston model basing on simulation results, and we compare the precision and efficiency of the two simulation methods.

\subsection{Crude Monte Carlo}
\label{sec:cMC}
We segment time $T$ to $n_T$ equal-length intervals via $T=n_Th$. For crude MC, we simulate the paths for $S(t)$ and $v(t)$ via
\begin{equation}\label{eqn:hat_S_cMC}
\hat{S}((i+1)h)=\hat{S}(ih)+r\hat{S}(ih)h+\sqrt{v(ih)}\hat{S}(ih)Z_{1i}
\end{equation}
\begin{equation}\label{eqn:hat_v_cMC}
\hat{v}((i+1)h)=\hat{v}(ih)-\lambda(\hat{v}(ih)-\bar{v})h+\eta\sqrt{\hat{v}(ih)}(\rho Z_{1i}+\sqrt{1-\rho^2}Z_{2i})
\end{equation}
for $i=0, \cdots, \frac{T}{h}-1$, where $\hat{S}(0)=S_0$, $\hat{v}(0)=v_0$, and $Z_{1i}, Z_{2i} \overset{\mathrm{iid}}{\sim}N(0,h), \forall i$.\\
Suppose $n_{\mathrm{P}}$ paths have been simulated. Then, the simulated call option price is calculated as
\begin{equation}\label{eqn:hat_c_cMC}
\hat{c}_{\mathrm{H}}=\frac{e^{-rT}}{n_{\mathrm{P}}}\sum_{j=1}^{n_{\mathrm{P}}}\left( \hat{S}_j(T)-K\right)^+
\end{equation}
where $\hat{S}_j(T)$ is the $j$th simulated $\hat{S}(T)$ among the $n_{\mathrm{P}}$ paths.

\subsection{Mixing Monte Carlo}
The idea of mixing method is to divide the evolution of underlying price, as given in Equation \eqref{eqn:hst_model}, into two independent degrees of freedom, one of which independent of the volatility process.\\
We have
\[\begin{aligned}
d(\ln{S(t)})&=\frac{1}{S(t)}dS(t)-\frac{1}{2S^2(t)}(dS(t))^2\\
&=(r-\frac{v(t)}{2})dt+\sqrt{v(t)}(\rho dW_1(t)+\sqrt{1-\rho^2}dW_2(t))\\
&=rdt+dY(t)-(1-\rho^2)\frac{v(t)}{2}dt+\sqrt{(1-\rho^2)v(t)}dW_2(t)\\
dv(t)&=-\lambda(v(t)-\bar{v})dt+\eta\sqrt{v(t)}dW_1(t)
\end{aligned}\]
where
\begin{equation}\label{eqn:mix_dY}
dY(t)=-\rho^2\frac{v(t)}{2}dt+\rho\sqrt{v(t)}dW_1(t)
\end{equation}
Then, for a given path implementation of $(v(t),Y(t))$, the effective initial underlying price $S_T^{\mathrm{eff}}$ (independent of $W_2$) and the effective volatility $\sigma_T^{\mathrm{eff}}$ is given by
\begin{equation}\label{eqn:mix_eff}
S_T^{\mathrm{eff}}=S_0e^{Y(T)}, \quad
\sigma_T^{\mathrm{eff}}=\sqrt{(1-\rho^2)\frac{\int_0^Tv(t)dt}{T}}
\end{equation}
With the same time segmentation setting as Subsection \ref{sec:cMC}, the implementation of mixing MC is via
\begin{equation}\label{eqn:hat_Y_mMC}
\hat{Y}((i+1)h)=\hat{Y}(ih)-\frac{\rho^2}{2}\hat{v}(ih)h+\rho\sqrt{\hat{v}(ih)}Z_i
\end{equation}
\begin{equation}\label{eqn:hat_v_mMC}
\hat{v}((i+1)h)=\hat{v}(ih)-\lambda(\hat{v}(ih)-\bar{v})h+\eta\sqrt{\hat{v}(ih)}Z_i
\end{equation}
for $i=0, \cdots, \frac{T}{h}-1$, where $\hat{Y}(0)=0$, $\hat{v}(0)=v_0$, and $Z_i\overset{\mathrm{iid}}{\sim}N(0,h), \forall i$.\\
Suppose $n_{\mathrm{P}}$ paths have been simulated. Then, the simulated call option price is calculated as
\begin{equation}\label{eqn:hat_c_mMC}
\hat{c}_{\mathrm{H}}=\frac{1}{n_{\mathrm{P}}}\sum_{j=1}^{n_{\mathrm{P}}}c_{\mathrm{BS}}(\hat{S}_{T,j}^{\mathrm{eff}}, K, r, T, \hat{\sigma}_{T,j}^{\mathrm{eff}})
\end{equation}
where
\begin{equation}\label{eqn:eff_hat}
\hat{S}_{T,j}^{\mathrm{eff}}=S_0e^{\hat{Y}_j(T)}, \quad
\hat{\sigma}_{T,j}^{\mathrm{eff}}=\sqrt{(1-\rho^2)\frac{h\sum_{i=0}^{T/h-1}\hat{v}_j(ih)}{T}}
\end{equation}
with $\hat{Y}_j(T)$ being the $j$th simulated $\hat{Y}(T)$ and $\hat{v}_j(ih)$ being the $j$th simulated $\hat{v}(ih)$ among the $n_{\mathrm{P}}$ paths, and $c_{\mathrm{BS}}(\cdot)$ is given in Equation \eqref{eqn:bs_call}.

\subsection{Simulation Results}
\label{sec:simu_result}
Consider the setting of Example 6.2.2 in \cite{glasserman2004monte}, where $S_0=K=100$, $v_0=\bar{v}=0.04$, $r=0.05$, $T=1$, $\lambda=1.2$, $\eta=0.3$, and $\rho=-0.5$. The theoretical call option price from Heston model, either according to \cite{glasserman2004monte} or calculated via Equation \eqref{eqn:hst_form}, is
$$c_{\mathrm{H}}=10.3009$$
We implement the crude MC (via Equation \eqref{eqn:hat_S_cMC} to \eqref{eqn:hat_c_cMC}) and mixing MC (via Equation \eqref{eqn:hat_Y_mMC} to \eqref{eqn:hat_c_mMC}) with $n_T=1000$ and multiple $n_{\mathrm{P}}$.

\begin{figure}[H]
\centering
\includegraphics[scale=0.8]{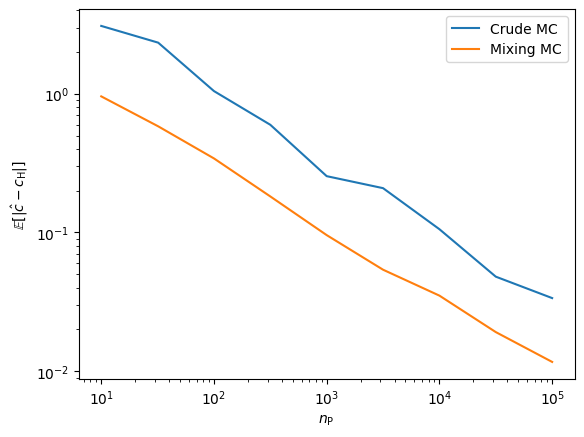}
\caption{Sample average of 50 replications of $|\hat{c}_{\mathrm{H}}-c_{\mathrm{H}}|$}
\label{fig:simu_accuracy}
\end{figure}

Figure \ref{fig:simu_accuracy} displays the sample average of 50 replications of $|\hat{c}-c_{\mathrm{H}}|$ with different values of $n_{\mathrm{P}}$. As the figure implies, first, the generated $\hat{c}$ from both MC methods converge to $c_{\mathrm{H}}$ as the number of simulated paths $n_{\mathrm{P}}$ increases, meaning that the Heston pricing method given in Equation \eqref{eqn:hst_form} is valid. Second, at each value of $n_{\mathrm{P}}$, the $\hat{c}$ generated from mixing MC is closer to $c_{\mathrm{H}}$ than that from crude MC, meaning that mixing MC has a higher accuracy under the same setting of $n_T$ and $n_{\mathrm{P}}$. In addition, the linear-like relationships between $\mathbb{E}[|\hat{c}-c_{\mathrm{H}}|]$ and $n_{\mathrm{P}}$ on the log scaled vertical and horizontal axes implies that mixing MC has a faster convergence rate than crude MC.

\begin{figure}[H]
\centering
\includegraphics[scale=0.6]{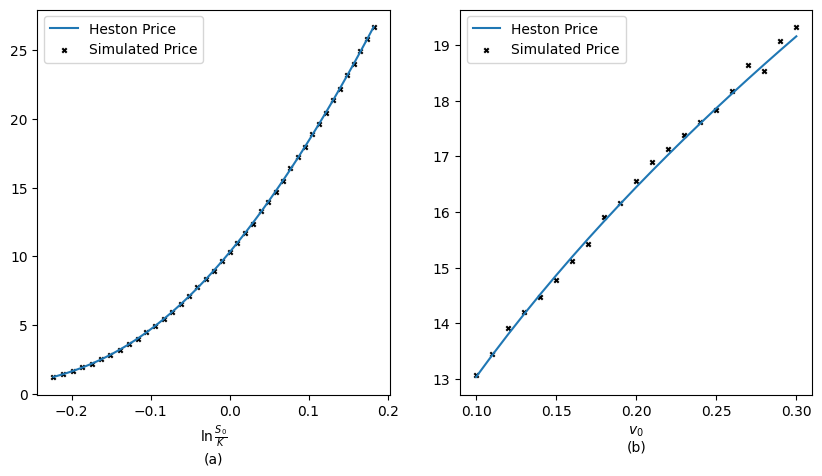}
\caption{Heston prices and simulated prices for various $S_0$ and $v_0$}
\label{fig:simu_mixing}
\end{figure}

Figure \ref{fig:simu_mixing} shows the Heston call option prices and simulated call option prices basing on mixing MC with $n_T=100$ and $n_{\mathrm{P}}=10000$ under the same parameter setting before except that $S_0$ varies from 80 to 120 in sub-figure (a) and that $v_0$ varies from 0.1 to 0.3 in sub-figure (b). The figure implies that call option prices generated from mixing MC method converge to Heston pricing for different parameter setting.

\newpage
\section{Model Testing}

\subsection{Put-Call Parity}
Continuing with the example in Section \ref{sec:simu_result}, we now verify that the put-call parity stated in Equation \eqref{eqn:parity} holds.

According to Equation \eqref{eqn:hst_form} and \eqref{eqn:P_j_put}, the put price of the example is
\[
p_{\mathrm{H}} = 5.4238.
\]
Therefore, we have
\[
\begin{aligned}
    c_{\mathrm{H}}-p_{\mathrm{H}}&=4.8770\\
    S_0-Ke^{-rT}&=4.8770
\end{aligned}
\]
meaning that the put-call parity holds for Heston model.

\subsection{Zero Strike}
Since we set $x=\ln{(S(t)e^{r\tau}/K}$ and $x$ is used in Equation \eqref{eqn:P_j_call}, we cannot set $K=0$ in calculating the analytical option price. By setting $K=0.001$, we have
\[
c_{\mathrm{H}}=99.9990
\]
which is close enough to $S_0=100$. Hence, we have verified that with strike $K=0$, the call option price will be equal to the underlying price.

\subsection{Varying $S_0$ and $v_0$}

As Figure \ref{fig:simu_mixing} implies, larger $S_0$ or larger $v_0$ results in higher call option price. This observation of Heston model is consistent with intuition, because larger $S_0$ means higher payoffs and higher probability of execution, and larger $v_0$ means higher volatility of the underlying and hence higher required return by the investors. Both scenarios will result in larger call option price.

\subsection{Deterministic Volatility}

\begin{figure}[H]
\centering
\includegraphics[scale=0.6]{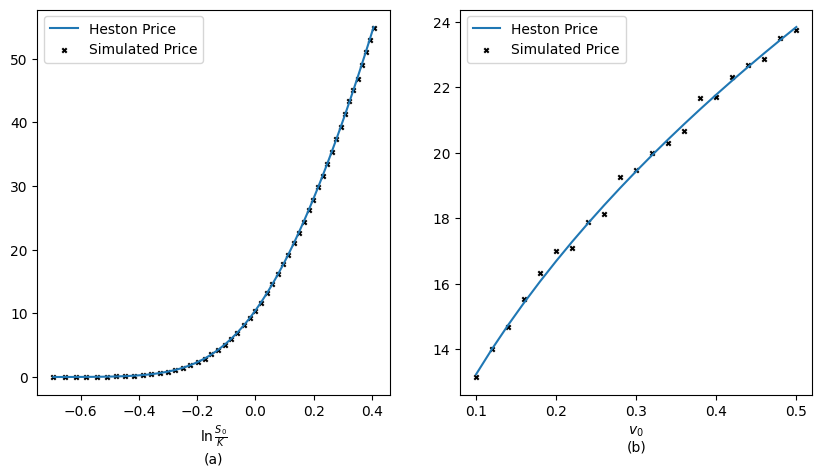}
\caption{Heston prices and simulated prices for various $S_0$ and $v_0$ with $eta=0$}
\label{fig:simu_eta0}
\end{figure}

Figure \ref{fig:simu_eta0} displays the Heston prices with $\eta=0$, calculated via Equation \eqref{eqn:sgm_eta0} and \eqref{eqn:bs_call}, and corresponding simulated prices for various $S_0$ and $v_0$. The plot implies that the derivation of Heston price under $\eta=0$ in Section \ref{sec:eta0} is valid.

\newpage
\section{Greeks and IV}
\label{sec:greeks_iv}
From this section on, we focus on European options underlying on futures. The future price is calculated by $F(\tau)=S(\tau)e^{r\tau}$.

\subsection{Greeks from Mixing MC}

In this section, we derive 5 Greeks of Heston model, which are Delta ($\Delta_{\mathrm{H}}$), Gamma ($\Gamma_{\mathrm{H}}$), Vega ($\nu_{\mathrm{H}}$), Theta ($\Theta_{\mathrm{H}}$), and Rho ($\rho_{\mathrm{H}}$), from the equations of mixing MC given by Equation \eqref{eqn:mix_dY} to \eqref{eqn:hat_c_mMC}.

\subsubsection*{Delta}
\begin{figure}[H]
\centering
\includegraphics[scale=0.6]{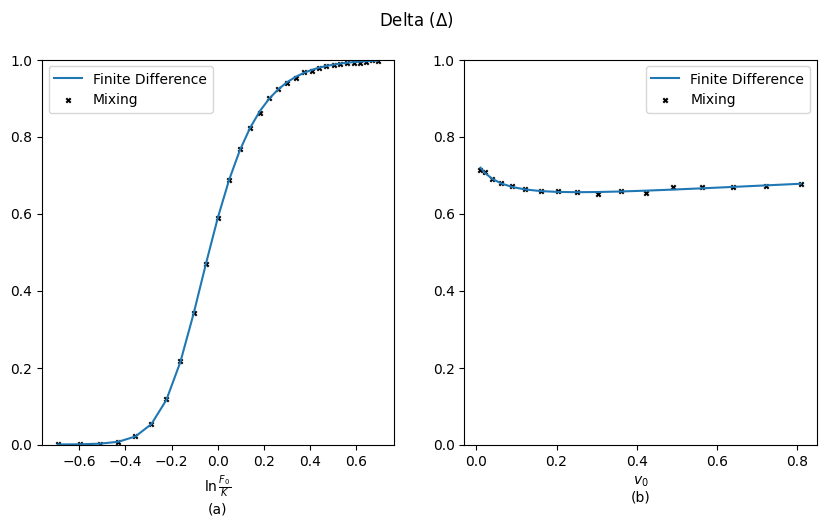}
\caption{Calculated Delta using mixing model (Equation \eqref{eqn:mix_delta}) and finite difference for various $F_0$ and $v_0$}
\label{fig:mixingDelta}
\end{figure}

Delta is the partial derivatives of option price with respect to initial future price, i.e.
$$\Delta_{\mathrm{H}}=\frac{\partial c_{\mathrm{H}}}{\partial F_0}.$$
Applying the chain rule to Equation \eqref{eqn:hat_c_mMC} and denoting $\hat{\theta}_{T,j}^{\mathrm{eff}}:=(\hat{F}_{T,j}^{\mathrm{eff}},K,r,T,\hat{\sigma}_{T,j}^{\mathrm{eff}})$, we get
\begin{equation}\label{eqn:mix_delta}
    \begin{aligned}\hat{\Delta}_{\mathrm{H}}&=\frac{1}{n_\mathrm{P}}\sum_{j=1}^{n_{\mathrm{P}}}\left(\Delta_{\mathrm{B}}(\hat{\theta}_{T,j}^{\mathrm{eff}})\frac{\partial\hat{F}_{T,j}^{\mathrm{eff}}}{\partial F_0}\right)\\ &=\frac{1}{n_\mathrm{P}}\sum_{j=1}^{n_{\mathrm{P}}}\left(\Delta_{\mathrm{B}}(\hat{\theta}_{T,j}^{\mathrm{eff}})e^{\hat{Y}_j(T)}\right)\end{aligned}
\end{equation}
where $\Delta_{\mathrm{B}}(\cdot)$ is the equation for Delta from Black model, calculated as
$$\Delta_{\mathrm{B}}(\theta)=e^{-rT}\Phi(d_{+})$$
Figure \ref{fig:mixingDelta} displays the calculated Delta varying $F_0$ and $v_0$ separately. Results support the validity of mixing method.

\subsubsection*{Gamma}
\begin{figure}[H]
\centering
\includegraphics[scale=0.6]{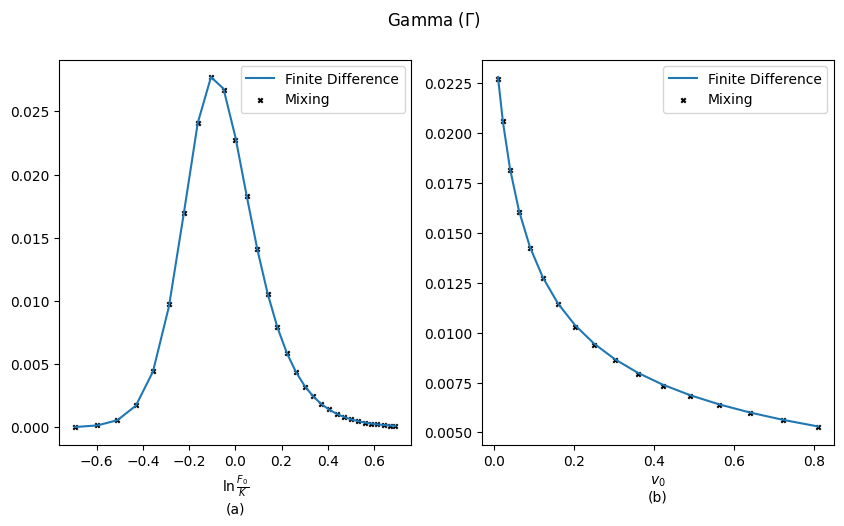}
\caption{Calculated Gamma using mixing model (Equation \eqref{eqn:mix_gamma}) and finite difference for various $F_0$ and $v_0$}
\label{fig:mixingGamma}
\end{figure}

Gamma is the second order partial derivatives of option price with respect to initial future price, i.e.,
$$\Gamma_{\mathrm{H}}=\frac{\partial^2 c_{\mathrm{H}}}{\partial F_0^2}.$$
Continuing from Equation \eqref{eqn:mix_delta}, we get
\begin{equation}\label{eqn:mix_gamma}
    \begin{aligned}\hat{\Gamma}_{\mathrm{H}}&=\frac{\partial\hat{\Delta}_{\mathrm{H}}}{\partial F_0}\\ &=\frac{1}{n_\mathrm{P}}\sum_{j=1}^{n_{\mathrm{P}}}\left(\Gamma_{\mathrm{B}}(\hat{\theta}_{T,j}^{\mathrm{eff}})e^{2\hat{Y}(T)}\right)\end{aligned}
\end{equation}
where $\Gamma_{\mathrm{B}}(\cdot)$ is the equation for Gamma from Black model, calculated as
$$\Gamma_{\mathrm{B}}(\theta)=\frac{\phi(d_{+})}{Fe^{-rT}\sigma\sqrt{T}}$$
with $\phi(x)=\frac{1}{\sqrt{2\pi}}e^{-\frac{x^2}{2}}$.
Figure \ref{fig:mixingGamma} displays the calculated Gamma varying $F_0$ and $v_0$ separately. Results support the validity of mixing method.

\subsubsection*{Vega}
\begin{figure}[H]
\centering
\includegraphics[scale=0.6]{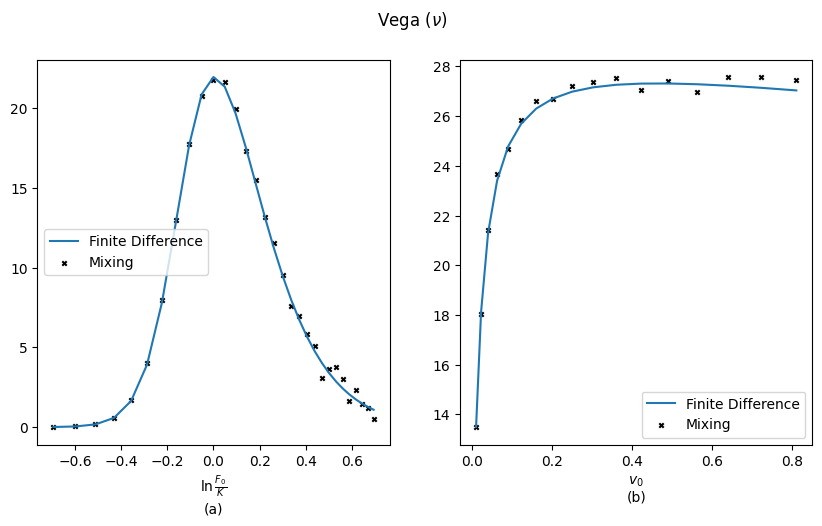}
\caption{Calculated Delta using mixing model (Equation \eqref{eqn:mix_vega}) and finite difference for various $F_0$ and $v_0$}
\label{fig:mixingVega}
\end{figure}

Vega is the partial derivatives of option price with respect to initial volatility, i.e.,
$$\nu_{\mathrm{H}}=\frac{\partial c_{\mathrm{H}}}{\partial\sqrt{v_0}}=2\sqrt{v_0}\frac{\partial c_{\mathrm{H}}}{\partial v_0}.$$
Applying the chain rule to Equation \eqref{eqn:hat_c_mMC}, we get
\begin{equation}\label{eqn:mix_vega}
    \begin{aligned}
        \hat{\nu}_{\mathrm{H}}&=\frac{2\sqrt{v_0}}{n_\mathrm{P}}\sum_{j=1}^{n_{\mathrm{P}}}\left(\Delta_{\mathrm{B}}(\hat{\theta}_{T,j}^{\mathrm{eff}})\frac{\partial\hat{F}_{T,j}^{\mathrm{eff}}}{\partial v_0}+\nu_{\mathrm{B}}(\hat{\theta}_{T,j}^{\mathrm{eff}})\frac{\partial\hat{\sigma}_{T,j}^{\mathrm{eff}}}{\partial v_0}\right)
    \end{aligned}
\end{equation}
where $\nu_{\mathrm{B}}(\cdot)$ is the equation for Vega from Black model, calculated as
\begin{equation}
    \nu_{\mathrm{B}}(\theta)=Fe^{-rT}\phi(d_+)\sqrt{T}.
\end{equation}
Then, according to Equation \eqref{eqn:eff_hat}, we have
\begin{equation}
    \frac{\partial\hat{F}_{T,j}^{\mathrm{eff}}}{\partial v_0}=\hat{F}_{T,j}^{\mathrm{eff}}\frac{\partial\hat{Y}_j(T)}{\partial v_0}
\end{equation}
where $\partial\hat{Y}_j(T)/\partial v_0$, according to Equation \eqref{eqn:hat_Y_mMC}, can be generated via iteration
\begin{equation}
    \frac{\partial\hat{Y}((i+1)h)}{\partial v_0}=\frac{\partial\hat{Y}(ih)}{\partial v_0}+\left(-\frac{\rho^2}{2}h+\frac{\rho Z_i}{2\sqrt{\hat{v}(ih)}}\right)\frac{\partial\hat{v}(ih)}{\partial v_0}, \quad \frac{\partial\hat{Y}(0)}{\partial v_0}=0.
\end{equation}
The only term unknown in the above equation is $\partial\hat{v}(ih)/\partial v_0$, which, according to Equation \eqref{eqn:hat_v_mMC}, can be generated via iteration
\begin{equation}
    \frac{\partial\hat{v}((i+1)h)}{\partial v_0}=\left(1-\lambda h+\frac{\eta Z_i}{2\sqrt{\hat{v}(ih)}}\right)\frac{\partial\hat{v}(ih)}{\partial v_0}, \quad \frac{\partial\hat{v}(0)}{\partial v_0}=1.
\end{equation}
In addition, according to \eqref{eqn:eff_hat} we have
\begin{equation}
    \frac{\partial\hat{\sigma}_{T}^{\mathrm{eff}}}{\partial v_0}=\frac{1}{2}\sqrt{\frac{(1-\rho^2)h}{T\sum_{i=0}^{T/h-1}\hat{v}(ih)}}\sum_{i=0}^{T/h-1}\frac{\partial\hat{v}(ih)}{\partial v_0}.
\end{equation}
Figure \ref{fig:mixingVega} displays the calculated Vega varying $F_0$ and $v_0$ separately. Results support the validity of mixing method. In addition, the variance of Vega generated from mixing model has higher variance when $F_0$ becomes larger.

\subsubsection*{Theta}

\begin{figure}[H]
\centering
\includegraphics[scale=0.6]{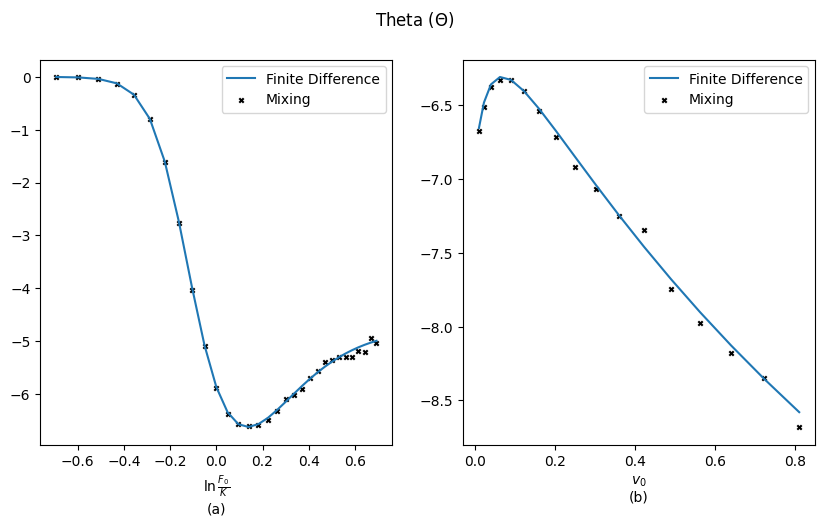}
\caption{Calculated Theta using mixing model (Equation \eqref{eqn:mix_theta}) and finite difference for various $F_0$ and $v_0$}
\label{fig:mixingTheta}
\end{figure}

Theta is the opposite of partial derivatives of option price with respect to time to maturity, i.e.,
\[
\Theta_{\mathrm{H}}=-\frac{\partial c_{\mathrm{H}}}{\partial{T}}.
\]
However, in practical implementation, due to the discretization of time, we compute theta as
\[
\hat{\Theta}_{\mathrm{H}}=-\frac{1}{n_T}\frac{\partial\hat{c}_{\mathrm{H}}}{\partial h}.
\]
Applying chain rule to Equation \eqref{eqn:hat_c_mMC}, we get
\begin{equation}\label{eqn:mix_theta}
    \begin{aligned}\hat{\Theta}_{\mathrm{H}}&=-\frac{1}{n_{\mathrm{P}}n_T}\sum_{j=1}^{n_{\mathrm{p}}}\left(\Delta_{\mathrm{B}}(\hat{\theta}_{T,j}^{\mathrm{eff}})\frac{\partial\hat{F}_{T,j}^{\mathrm{eff}}}{\partial h}+\nu_{\mathrm{B}}(\hat{\theta}_{T,j}^{\mathrm{eff}})\frac{\partial\hat{\sigma}_{T,j}^{\mathrm{eff}}}{\partial h}+\Theta_{\mathrm{B}}(\hat{\theta}_{T,j}^{\mathrm{eff}})\right)\\ &=-\frac{1}{n_{\mathrm{P}}n_T}\sum_{j=1}^{n_{\mathrm{p}}}\left(\Delta_{\mathrm{B}}(\hat{\theta}_{T,j}^{\mathrm{eff}})\hat{F}_{T,j}^{\mathrm{eff}}\frac{\partial\hat{Y}_j(T)}{\partial h}+\nu_{\mathrm{B}}(\hat{\theta}_{T,j}^{\mathrm{eff}})\frac{\partial\hat{\sigma}_{T,j}^{\mathrm{eff}}}{\partial h}+\Theta_{\mathrm{B}}(\hat{\theta}_{T,j}^{\mathrm{eff}})\right)\end{aligned}
\end{equation}
where $\Theta_{\mathrm{B}}(\cdot)$ is the equation for Theta from Black model, calculated as
\begin{equation}
    \Theta_{\mathrm{B}}(\theta)=-e^{-rT}\left(F\phi(d_+)\frac{\sigma}{2\sqrt{T}}+rK\Phi(d_-)\right).
\end{equation}
Notice that the evolution of $\hat{Y}$ and $\hat{v}$ in Equation \eqref{eqn:hat_Y_mMC} and \eqref{eqn:hat_v_mMC} contains a random variable $Z_i\overset{\mathrm{iid}}{\sim}N(0,h)$, which depends on $h$. Hence, we write it as $Z_i:=\sqrt{h}Z_i'$ with $Z_i'\overset{iid}{\sim}N(0,1)$. Then, $\partial\hat{Y}_j(T)/\partial h$, according to Equation \eqref{eqn:hat_Y_mMC}, can be generated via iteration
\begin{equation}
    \begin{aligned}\frac{\partial\hat{Y}((i+1)h)}{\partial h}&=\frac{\partial\hat{Y}(ih)}{\partial h}-\frac{\rho^2}{2}\frac{\partial(v(ih)h)}{\partial h}+\rho Z_i'\frac{\partial\sqrt{v(ih)h}}{\partial h}\\ &= \frac{\partial\hat{Y}(ih)}{\partial h}-\frac{1}{2}\left(\rho^2h-\frac{\rho Z_i}{\sqrt{v(ih)}}\right)\left(\frac{v(ih)}{h}+\frac{\partial v(ih)}{\partial h}\right)\end{aligned}
\end{equation}
where $\partial\hat{v}(ih)/\partial h$ can be generated via iteration
\begin{equation}
    \begin{aligned}\frac{\partial\hat{v}((i+1)h)}{\partial h}&=\frac{\partial\hat{v}(ih)}{\partial h}-\lambda\frac{\partial((\hat{v}(ih)-\bar{v})h)}{\partial h}+\eta Z_i'\frac{\partial\sqrt{\hat{v}(ih)h}}{\partial h}\\ &=(1-\lambda h)\frac{\partial v(ih)}{\partial h}-\lambda(v(ih)-\bar{v})+\frac{\eta Z_i}{2\sqrt{v(ih)}}\left(\frac{v(ih)}{h}+\frac{\partial v(ih)}{\partial h}\right)\end{aligned}.
\end{equation}
In addition, according to Equation \eqref{eqn:eff_hat} we have
\begin{equation}
    \begin{aligned}\frac{\partial\hat{\sigma}_T^{\mathrm{eff}}}{\partial h}&=\frac{\partial}{\partial h}\sqrt{(1-\rho^2)\frac{\sum_{i=0}^{n_T-1}\hat{v}(ih)}{n_T}}\\ &=\frac{1}{2}\sqrt{\frac{1-\rho^2}{n_T\sum_{i=1}^{n_T-1}\hat{v}(ih)}}\sum_{i=1}^{n_T-1}\frac{\partial\hat{v}(ih)}{\partial h}\\ &= \frac{1}{2}\sqrt{\frac{h(1-\rho^2)}{T\sum_{i=1}^{T/h-1}\hat{v}(ih)}}\sum_{i=1}^{T/h-1}\frac{\partial\hat{v}(ih)}{\partial h}\end{aligned}
\end{equation}
where we transform $h/T$ to $1/n_T$ because $n_T$ is independent of $h$.
Figure \ref{fig:mixingTheta} displays the calculated Theta varying $F_0$ and $v_0$ separately. Results support the validity of mixing method.

\subsubsection*{Rho}
\begin{figure}[H]
\centering
\includegraphics[scale=0.6]{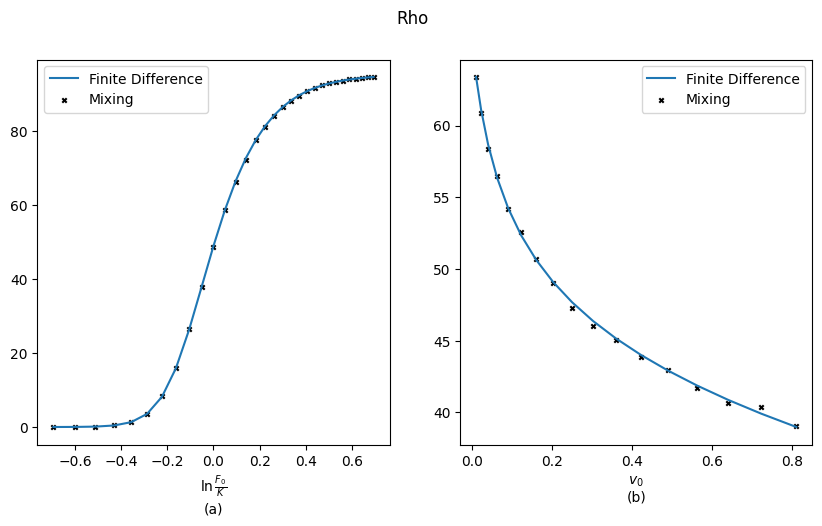}
\caption{Calculated Delta using mixing model (Equation \eqref{eqn:mix_rho}) and finite difference for various $F_0$ and $v_0$}
\label{fig:mixingRho}
\end{figure}

Rho is the partial derivatives of option price with respect to interest rate, i.e.,
\[
\rho_{\mathrm{H}}=\frac{\partial c_{\mathrm{H}}}{\partial r}.
\]
Since neither $\hat{S}_T^{\mathrm{eff}}$ nor $\hat{\sigma}_T^{\mathrm{eff}}$ depends on $r$, we have
\begin{equation}\label{eqn:mix_rho}
    \hat{\rho}_{\mathrm{H}}=\frac{1}{n_{\mathrm{P}}}\sum_{j=1}^{n_{\mathrm{P}}}\rho_{\mathrm{B}}(\hat{\theta}_{T,j}^{\mathrm{eff}})
\end{equation}
where $\rho_{\mathrm{B}}(\cdot)$ is the equation for Rho from Black model, calculated as
\begin{equation}
    \rho_{\mathrm{B}}(\theta)=KTe^{rT}\Phi(d_-).
\end{equation}
Figure \ref{fig:mixingRho} displays the calculated Rho varying $F_0$ and $v_0$ separately. Results support the validity of mixing method.

\subsection{IV Curves}
\label{sec:IV_curves}

IV is defined as the volatility that, when input into the Black-Scholes formula, results in an option price which is equal to the market price.  The Heston model captures the market phenomenon of IV smile, where IV is a convex curve with respect to the strike price $K$.

% One of the advantages of Heston model over Black-Scholes model is that it captures the market phenomenon of IV smile, where prices of options with strike closer to the underlying price (i.e., more at-the-money) imply a smaller volatility than those with strike further away from the underlying price (i.e., more in-the-money or out-of-the-money). 

In this section, we will visualize the IV curves from Heston model and analyze the relationships between IV curves and the parameters ($\rho$, $\eta$, $\lambda$, $v_0$, and $\bar{v}$) of Heston model. We continue with the same parameters setting as Section \ref{sec:simu_result}, except that we vary $K$ from $50$ to $200$ to generate one IV curve. Then, we vary one of the five parameters listed above while fixing the other four to investigate its effect on the IV curve.

\subsubsection*{Varying $\rho$}

\begin{figure}[H]
\centering
\includegraphics[scale=0.6]{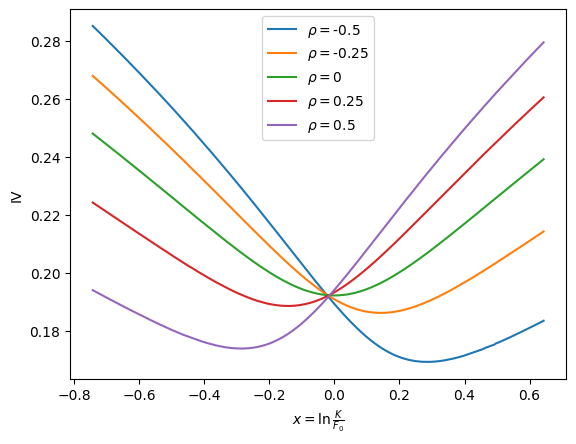}
\caption{IV curves with different $\rho$}
\label{fig:smile_rho}
\end{figure}

Figure \ref{fig:smile_rho} displays the IV curves with different $\rho$ ranging from $-0.5$ to $0.5$. The figure implies that the lowest point of the IV curve (i.e., the symmetric point) is achieved at $K>F_0$ when $\rho$ is negative (positive). Therefore, $\rho$ determines which $K$ is lowest IV (or the symmetry axis of IV curve) located.

\subsubsection*{Varying $\eta$ or $\lambda$}

\begin{figure}[H]
\centering
\includegraphics[scale=0.6]{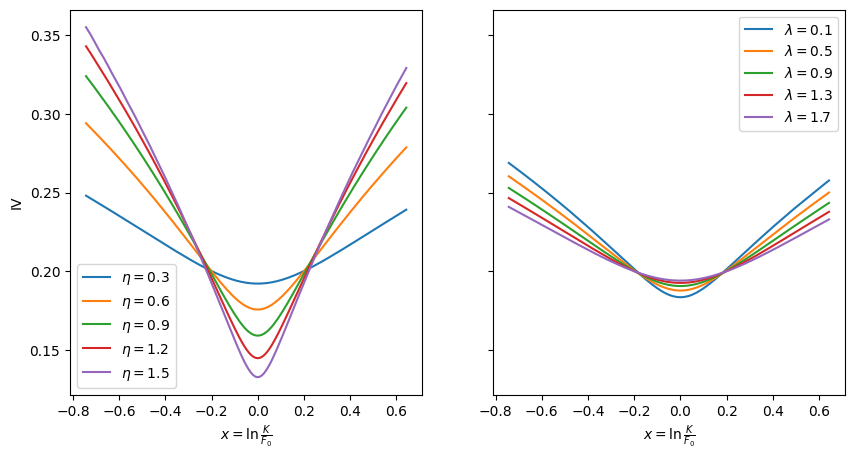}
\caption{IV curves with different $\eta$ or $\lambda$}
\label{fig:smile_lambda_eta}
\end{figure}

Figure \ref{fig:smile_lambda_eta} displays the IV curves with different $\eta$ ranging from $0.3$ to $1.5$ or different $\lambda$ ranging from $0.1$ to $1.7$. The figure implies that larger $\eta$ or $\lambda$ results in larger curvature of the IV curve. In addition, the curvature is much more sensitive to $\eta$ than to $\lambda$. Therefore, $\eta$ and $\lambda$ determine the curvature of IV curves, and the effect of $\eta$ is much more significant than that of $\lambda$.

\subsubsection*{Varying $v_0$ or $\bar{v}$}

\begin{figure}[H]
\centering
\includegraphics[scale=0.6]{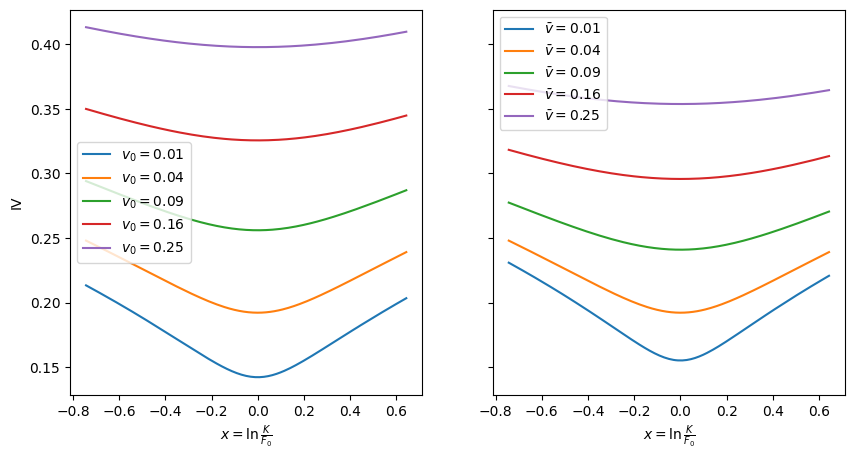}
\caption{IV curves with different $v_0$ or $\bar{v}$}
\label{fig:smile_v0_vbar}
\end{figure}

Figure \ref{fig:smile_v0_vbar} displays the IV curves with different $v_0$ or $\bar{v}$ ranging from $0.01$ to $0.25$. The figure implies that larger $v_0$ or $\bar{v}$ results in higher overall level and smaller curvature of the IV curve. In addition, the overall IV level is more sensitive to $v_0$ than to $\bar{v}$. Therefore, $v_0$ and $\bar{v}$ mainly determine the overall level of IV curves, and the effect of $v_0$ is slightly more significant than that of $\bar{v}$.

In section \ref{sec:calib}, we demonstrate an empirical validation method of Heston model using Crude Oil WTI data to calibrate. We apply the results of the above IV smile curves to tune the parameters to optimize the fitting of the curves. In addition, the curvature,  skewness, and kurtosis of each contract's features for selected dates.

\newpage
\section{Calibration}
\label{sec:calib}

In the following sections, we introduce a calibration example of extending the numerical analysis of Heston model through Crude Oil WTI data. 

We obtain the WTI futures and options data from \url{Barchart.com}. 

Before processing the data and developing corresponding loss functions and Gradient Descent Machine Learning models to optimize the choices of the five Heston constants, we first determine the values of number of days of a year $T = 365$ and the risk free interest rate $r = 0.036$. 

The study and experiments through the historical data obtained from the website support our selections of $T$ and $r$, through the Black-Scholes model.

In addition, we treat the collected WTI data like European-style contracts for simplicity.

Because of the complex nature of the hardship of acquiring historical implied volatility and option prices for the WTI contracts, we manually gather data from the website for 6 days: 04 24 2024, 04 25 2024, 04 26 2024, 04 30 2024, 05 09 2024, and 05 10 2024.

We provide brief pseudo codes for the 3 primary functions of our optimization algorithm in the following subsections.

\subsection{Data Processing}

\begin{verbatim}
1. Read data from CSV file.
2. Drop rows with missing values.
3. Convert `Strike' and `IV' columns to numeric, 
   stripping non-numeric characters.
4. Filter out rows where IV is very close to zero (either +0.01 or -0.01).
5. Group data by `Strike' and calculate the mean of `IV' for each group.
6. Reset index and rename columns appropriately.
7. Calculate the log of strike prices relative to close price.
8. Plot to visualize the relationship between transformed strike and IV.
\end{verbatim}

\subsection{Loss Function}
\begin{verbatim}
1. Define the loss function to calculate 
   the mean squared error between actual IVs and IVs predicted 
   by the Heston model using the current parameter set.
2. Parameters include volatility (sgm), long-run variance (vbar), 
   speed of mean reversion (lamb), vol of vol (eta), and correlation (rho).
3. Use the Heston model to estimate prices for given strike prices.
4. Convert these prices to implied volatilities.
5. Compute the mean squared error against actual IVs.
\end{verbatim}

\subsection{Gradient Descent Function}
\begin{verbatim}
1. Initialize parameters for the model.
2. For a predefined number of iterations:
   a. Perturb each parameter slightly upward and downward.
   b. Calculate the loss for each perturbed set of parameters.
   c. Estimate the gradient with respect to each parameter.
   d. Adjust the learning rate based on the current iteration.
   e. Update each parameter in the direction that reduces the loss.
   f. Print the current parameters and loss for monitoring.
3. After iterations, return the optimized parameters.
4. Use these parameters to calculate IVs and compare with actual IVs.
\end{verbatim}

During the gradient descent process, we set several restrictions on the parameters:

\[
0.02 \leq v_0 \leq 0.12
\]

\[
\max(0.02,v_0-0.04) \leq v_{bar} \leq \max(0.02,v_0+0.04)
\]

(in later section we set this to be a constant $0.073$)

\[
-2 \leq \lambda \leq 2
\]
(in later section we set this to be a constant $0.528$)

\[
0 \leq \eta \leq 2
\]

\[
-0.5 \leq \rho \leq 0.5
\]

The choices of the parameter bounds are based on numerious experiments and observations.

After these 3 sections, we plot both sets of IVs against the transformed strike prices to evaluate the model fit.

\subsection{Calibration Results}

\subsubsection{Fixing 0 Parameters}

We first regard all of the 5 parameters as variables and use gradient descent to optimize the mean square error of market and calibrated IVs.

\begin{figure}[H]
\centering
\includegraphics[scale=0.33]{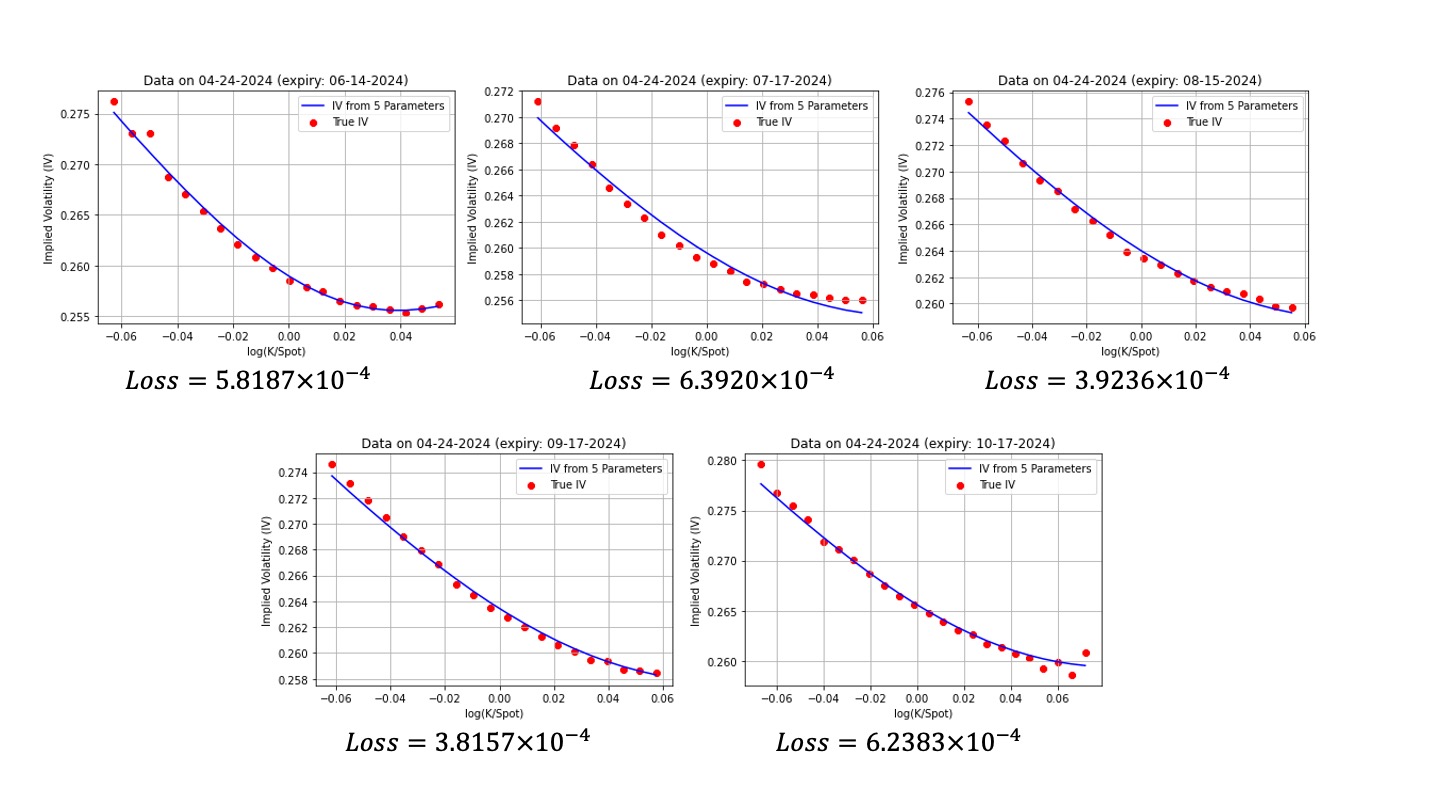}
\caption{True and calibrated IV for contracts of 5 expiry dates with loss values of data \\ on 04-24-2024, fixing 0 Parameters.}
\label{fig:calib_0424_fix0}
\end{figure}

\begin{figure}[H]
\centering
\includegraphics[scale=0.33]{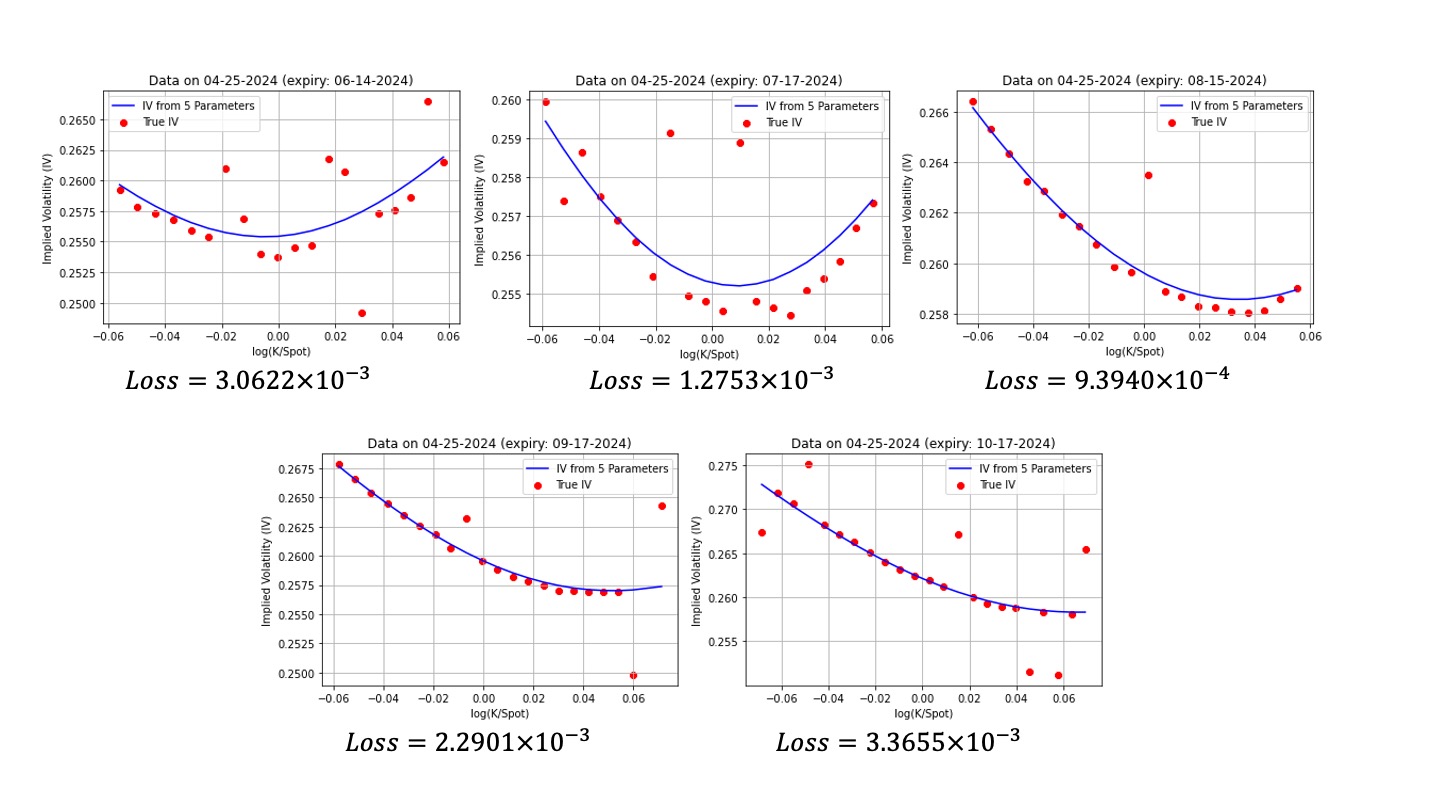}
\caption{True and calibrated IV for contracts of 5 expiry dates with loss values of data on 04-25-2024, fixing 0 Parameters.}
\label{fig:calib_0425_fix0}
\end{figure}

\begin{figure}[H]
\centering
\includegraphics[scale=0.33]{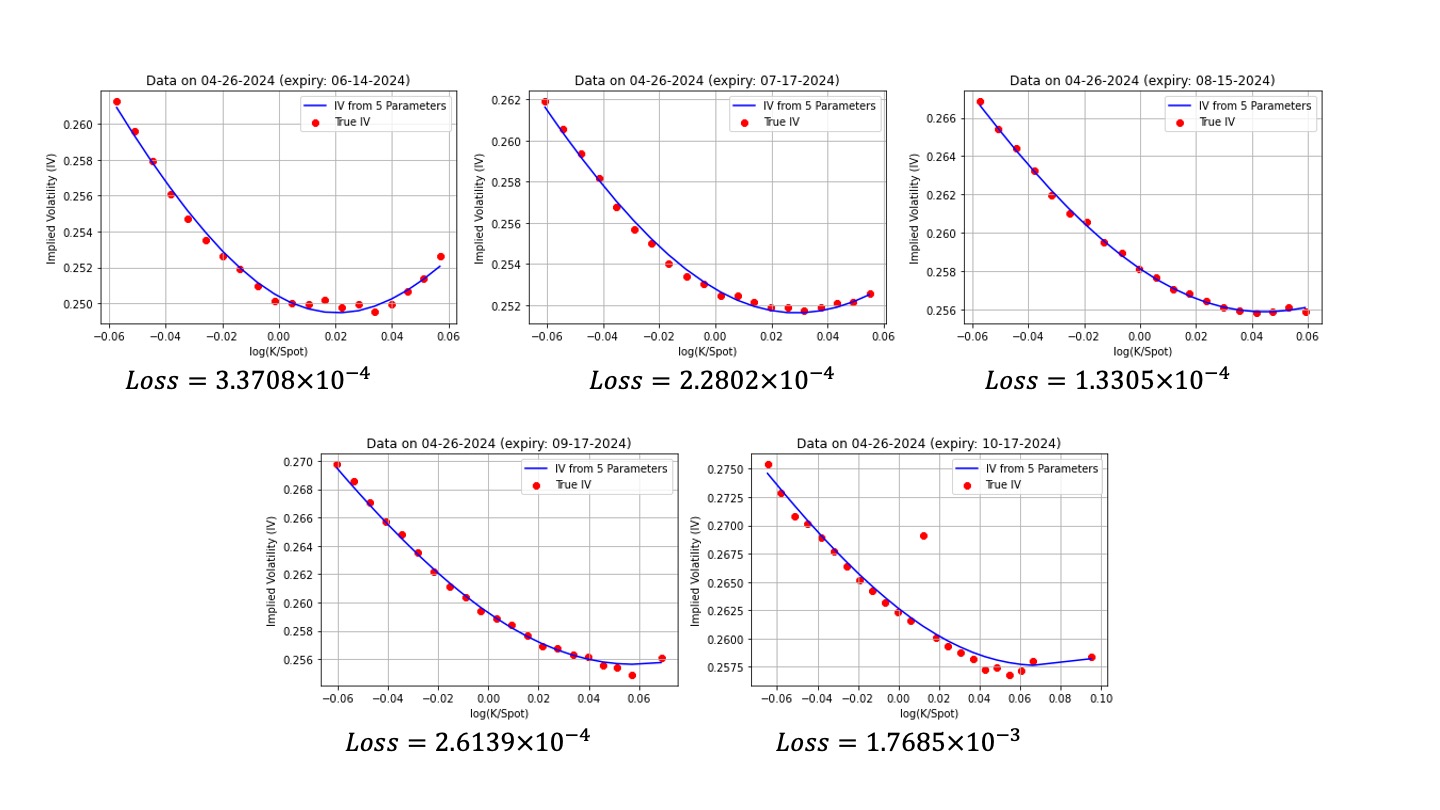}
\caption{True and calibrated IV for contracts of 5 expiry dates with loss values of data on 04-26-2024, fixing 0 Parameters.}
\label{fig:calib_0426_fix0}
\end{figure}

Figure \ref{fig:calib_0424_fix0} to \ref{fig:calib_0426_fix0} display the true and fitted IVs of option contracts of 5 expiry dates (ranging from 06-14 to 10-17 in 2024) on  3 days (04-24 to 04-26 in 2024), correspondingly. 

Table \ref{tab:calib_fix0} shows the calibrated 5 parameters of the 5 contracts on 3 days. 
\begin{table}[H]
    \centering
    \begin{tabular}{|l||*{5}{c|}}
    \hline
    \diagbox[width=6em]{Date}{Expiry} &\makebox[3.5em]{06-14}  &\makebox[3.5em]{07-17} & \makebox[3.5em]{08-15} & \makebox[3.5em]{09-17} & \makebox[3.5em]{10-17} \\\hline\hline
    04-24 & \begin{tabular}{@{}c@{}c@{}}$v_0=0.0755$ \\ $\bar{v}=0.0748$ \\ $\lambda=0.4533$ \\ $\eta=0.8996$ \\ $\rho=-0.1679$\end{tabular} & \begin{tabular}{@{}c@{}c@{}}$v_0=0.0736$ \\ $\bar{v}=0.0741$ \\ $\lambda=0.4545$ \\ $\eta=0.6050$ \\ $\rho=-0.1997$\end{tabular} & \begin{tabular}{@{}c@{}c@{}}$v_0=0.0777$ \\ $\bar{v}=0.0770$ \\ $\lambda=0.4506$ \\ $\eta=0.6020$ \\ $\rho=-0.1994$\end{tabular} & \begin{tabular}{@{}c@{}c@{}}$v_0=0.0791$ \\ $\bar{v}=0.0790$ \\ $\lambda=0.4495$ \\ $\eta=0.6030$ \\ $\rho=-0.2046$\end{tabular} & \begin{tabular}{@{}c@{}c@{}}$v_0=0.0835$ \\ $\bar{v}=0.0830$ \\ $\lambda=0.4494$ \\ $\eta=0.6530$ \\ $\rho=-0.2071$\end{tabular} \\\hline\hline
    04-25 & \begin{tabular}{@{}c@{}c@{}}$v_0=0.0720$ \\ $\bar{v}=0.0721$ \\ $\lambda=0.4501$ \\ $\eta=0.8019$ \\ $\rho=-0.0434$\end{tabular} & \begin{tabular}{@{}c@{}c@{}}$v_0=0.0715$ \\ $\bar{v}=0.0714$ \\ $\lambda=0.4556$ \\ $\eta=0.6117$ \\ $\rho=-0.0018$\end{tabular} & \begin{tabular}{@{}c@{}c@{}}$v_0=0.0752$ \\ $\bar{v}=0.0750$ \\ $\lambda=0.4505$ \\ $\eta=0.5993$ \\ $\rho=-0.0718$\end{tabular} & \begin{tabular}{@{}c@{}c@{}}$v_0=0.0781$ \\ $\bar{v}=0.0780$ \\ $\lambda=0.4515$ \\ $\eta=0.6416$ \\ $\rho=-0.1220$\end{tabular} & \begin{tabular}{@{}c@{}c@{}}$v_0=0.0810$ \\ $\bar{v}=0.0810$ \\ $\lambda=0.4511$ \\ $\eta=0.6405$ \\ $\rho=-0.1528$\end{tabular} \\\hline\hline
    04-26 & \begin{tabular}{@{}c@{}c@{}}$v_0=0.0700$ \\ $\bar{v}=0.0700$ \\ $\lambda=0.4503$ \\ $\eta=0.8498$ \\ $\rho=-0.0779$\end{tabular} & \begin{tabular}{@{}c@{}c@{}}$v_0=0.0719$ \\ $\bar{v}=0.0722$ \\ $\lambda=0.4515$ \\ $\eta=0.7050$ \\ $\rho=-0.0824$\end{tabular} & \begin{tabular}{@{}c@{}c@{}}$v_0=0.0761$ \\ $\bar{v}=0.0760$ \\ $\lambda=0.4502$ \\ $\eta=0.6698$ \\ $\rho=-0.1193$\end{tabular} & \begin{tabular}{@{}c@{}c@{}}$v_0=0.0791$ \\ $\bar{v}=0.0790$ \\ $\lambda=0.4499$ \\ $\eta=0.6810$ \\ $\rho=-0.1530$\end{tabular} & \begin{tabular}{@{}c@{}c@{}}$v_0=0.0834$ \\ $\bar{v}=0.0830$ \\ $\lambda=0.4503$ \\ $\eta=0.7008$ \\ $\rho=-0.1812$\end{tabular} \\\hline
    \end{tabular}
    \caption{Calibrated parameters of contracts of 5 expiry dates on 3 days (all dates in year 2024), fixing 0 parameters.}
    \label{tab:calib_fix0}
\end{table}

Basing on these results, we have the following observations. First, parameters $v_0$ and $\bar{v}$ are relatively stable among the same option contract on different days, but they vary a lot on contracts with different expiry dates. More specifically, as time to maturity increases, both $v_0$ and $\bar{v}$ generally increase. This implies that traders generally embed higher volatility levels to contracts with longer time to maturity. In addition, we also observe that the difference between $v_0$ and $\bar{v}$ in each block in Table \ref{tab:calib_fix0} is very small. This is because we use the same value for $v_0$ and $\bar{v}$ as initial parameters, and as shown in Section \ref{sec:IV_curves}, the effects of $v_0$ and $\bar{v}$ on IV curves are very similar. Therefore, the change in these two parameters behave similarly through the gradient descent method.

First, the parameter $\lambda$ are very stable and constant among all contracts of different expiry dates and all days. This is consistent with the result in Section \ref{sec:IV_curves}, which shows that although $\lambda$ affect the curvature of IV curves, the effect is much smaller than that of $\lambda$. Since we used the same $\lambda$ value (which is $0.45$) as initial parameters, the output $\lambda$ does not change much.

Second, the parameter $\eta$ is relatively unstable among the same contract on different days. This implies that the volatility of volatility embedded in the market is not constant. Therefore, the assumption of constant volatility of volatility in Heston model may be one of its limitations.

Third, the parameter $\rho$ is unstable and does not have any pattern among any option contract on different days. However, there is an downward trends on $\rho$ when time to maturity increases. This also reflects a limitation of Heston model which assumes constant correlation between underlying price and volatility, and results imply that the correlation becomes more negative when time to maturity increases.

Notice that contracts with different expiry dates are considered as different commodity assets with different term structures. How parameters changes depends on the maturity of the contract. for example, ATM volatility decreases for longer maturities. The problem with Heston is that the parameters are not intuitive, for example ATM volatility  (which changes each day) is not a parameter but a complex function of the parameters; or skew, defined as a slope.

\subsubsection{Fix all 5 Parameters}

Following the dynamics of the Heston model for option pricing, it is nature to choose all 5 parameters to be constant for the lifetime of a particular contract. To test this we compute the average of the parameters presented in the table above and compute the loss and plot the figure as below for an example day of April 26, 2024.

\begin{table}[H]
    \[
    \begin{array}{|c|c|c|c|c|c|}
    \hline
    \text{Date} & v_0 & \bar{v} & \lambda & \eta & \rho \\
    \hline
    04/24/2024 & 0.07375538 & 0.07669798 & 0.50016636 & 0.6269777 & -0.16326973 \\
    04/25/2024 & 0.07300135 & 0.07298108 & 0.45087972 & 0.69693631 & -0.01901624 \\
    04/26/2024 & 0.07092082 & 0.07021198 & 0.4664877 & 0.60722274 & -0.14697237 \\
    04/30/2024 & 0.0694863 & 0.07514112 & 0.7130847 & 0.7205338 & -0.48566705 \\
    05/09/2024 & 0.06774554 & 0.06728420 & 0.49148604 & 0.75141233 & -0.03003124 \\
    05/10/2024 & 0.06635441 & 0.07734824 & 0.5454521 & 0.64249337 & -0.21371213 \\
    \hline
    \text{Mean} & 0.07021063 & 0.07327743 & 0.5279261 & 0.67426271 & -0.17644479 \\
    \hline
    \end{array}
    \]
    \caption{Calibrated example, fixing all 5 parameters}
    \label{tab:calib_fix5}
\end{table}

This last row provides the average values of the parameters across all the dates.

\begin{figure}[H]
\centering
\includegraphics[scale=0.3]{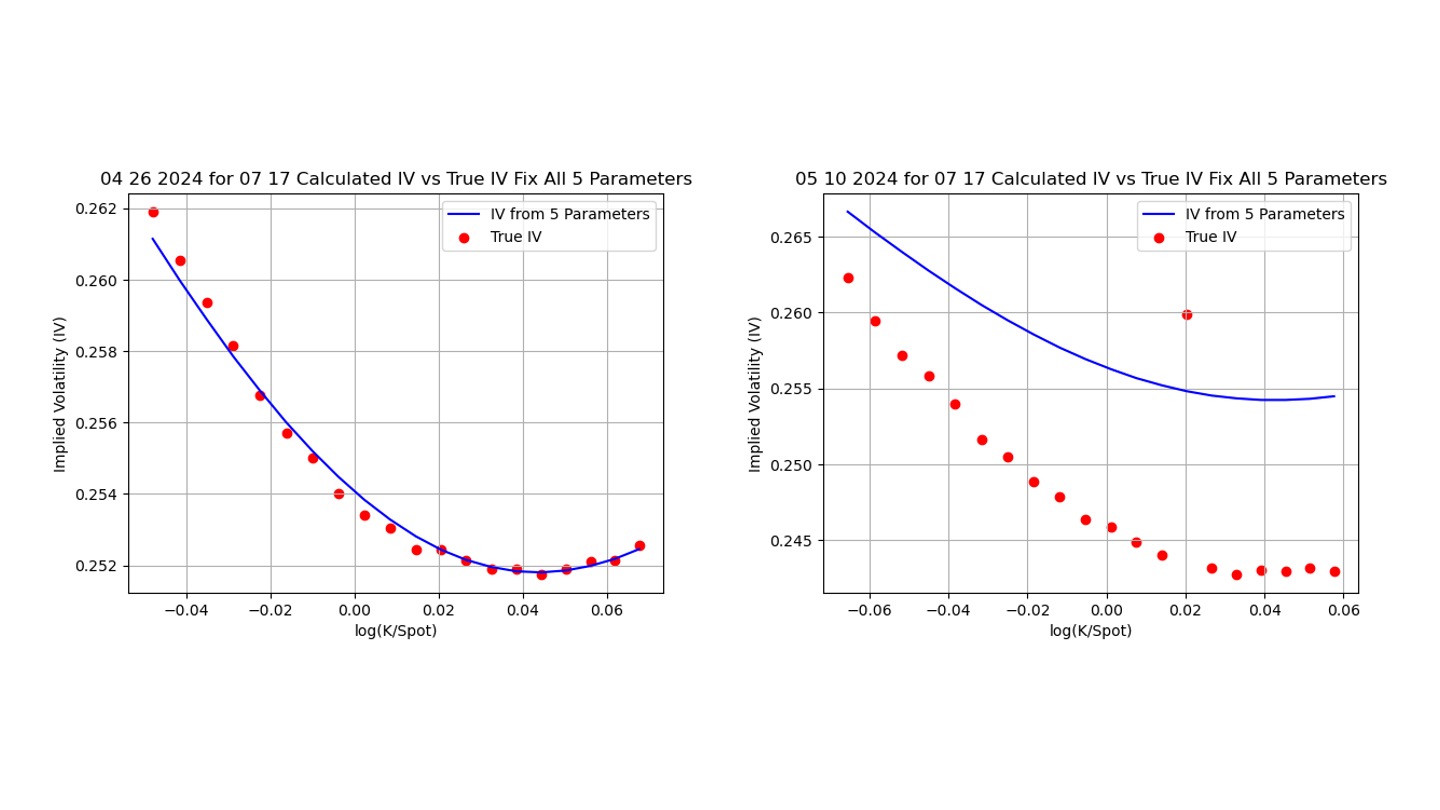}
\caption{True and calibrated IV for two days, fixing 5 Parameters (using the mean)}
\label{fig:07_17_Fix_5}
\end{figure}

For the Crude Oil WTI contract expired on July 17, 2024, we test with these 5 parameters on two dates: April 26 2024 and May 10 2024. Figure \ref{fig:07_17_Fix_5} displays the fitting result, where the fitting loss of 04-26 is $3.1686\times10^{-4}$ and that of 05-10 is $9.5305\times10^{-3}$.

Basing on their mean square losses and plot visualizations, fitting of 04-26 seems to outperform 05-10 greatly. One reasonable explanation is 05-10 data contains one outlier, for which the fix all 5 parameters method seem to fail for this particular date of the contract.

Therefore we propose fixing 2 of the 5 parameters to balance the model dynamics and the real market complexity.

\subsubsection{Fixing 2 parameters}

\begin{figure}[H]
\centering
\includegraphics[scale=0.33]{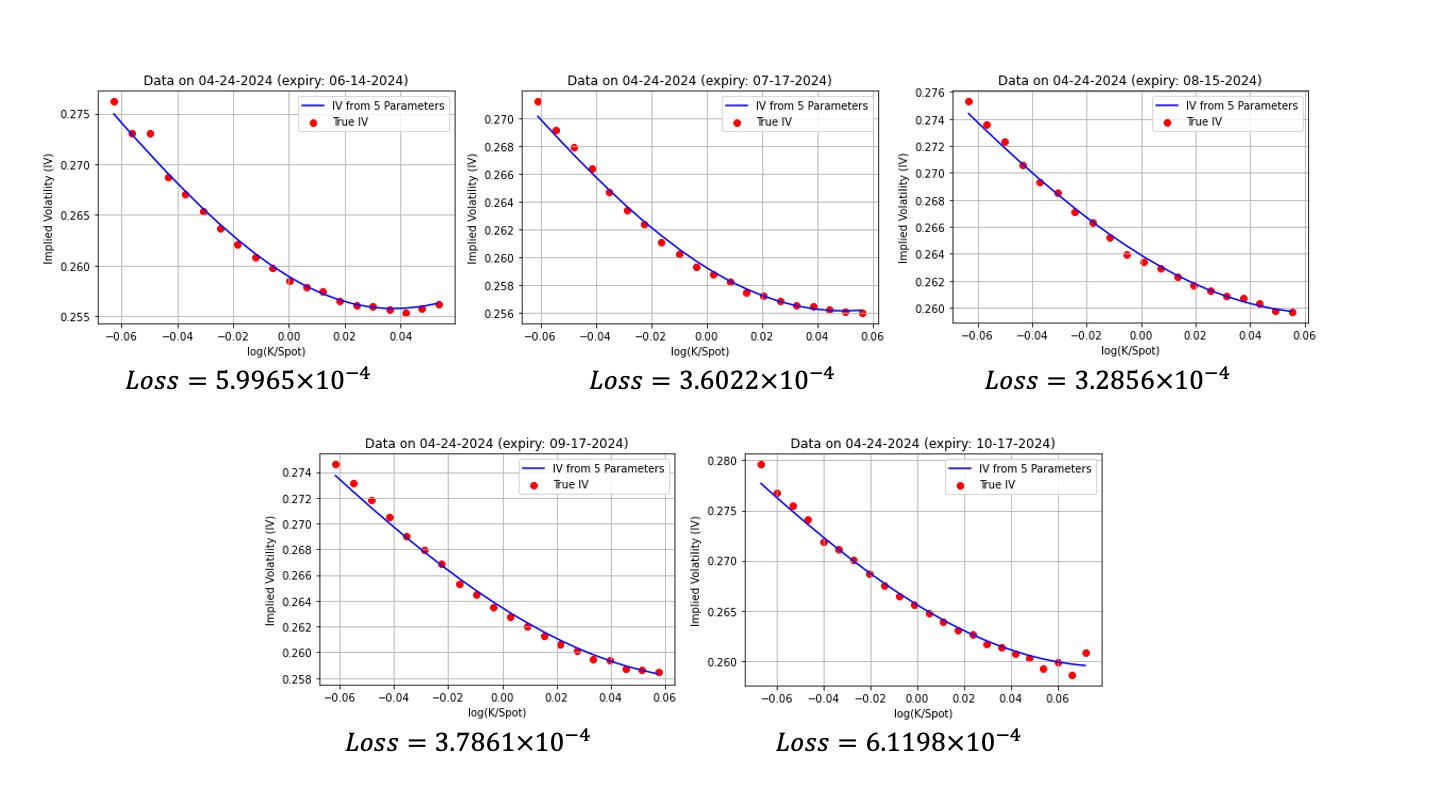}
\caption{True and calibrated IV for contracts of 5 expiry dates with loss values of data on 04-24-2024, fixing $\bar{v}=0.0763$ and $\lambda=0.45$.}
\label{fig:calib_0424_fix2}
\end{figure}

\begin{figure}[H]
\centering
\includegraphics[scale=0.33]{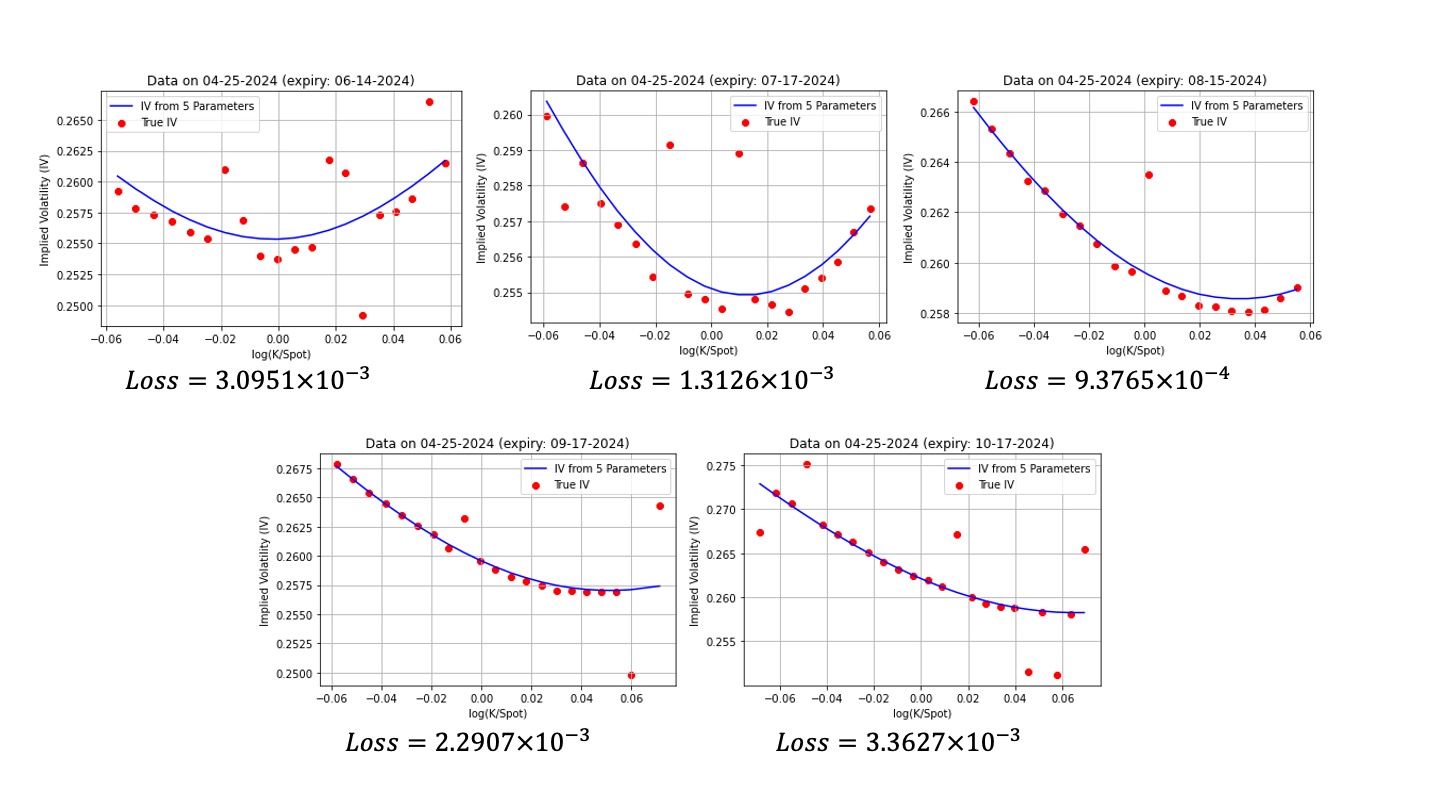}
\caption{True and calibrated IV for contracts of 5 expiry dates with loss values of data on 04-25-2024, fixing $\bar{v}=0.0763$ and $\lambda=0.45$.}
\label{fig:calib_0425_fix2}
\end{figure}

\begin{figure}[H]
\centering
\includegraphics[scale=0.33]{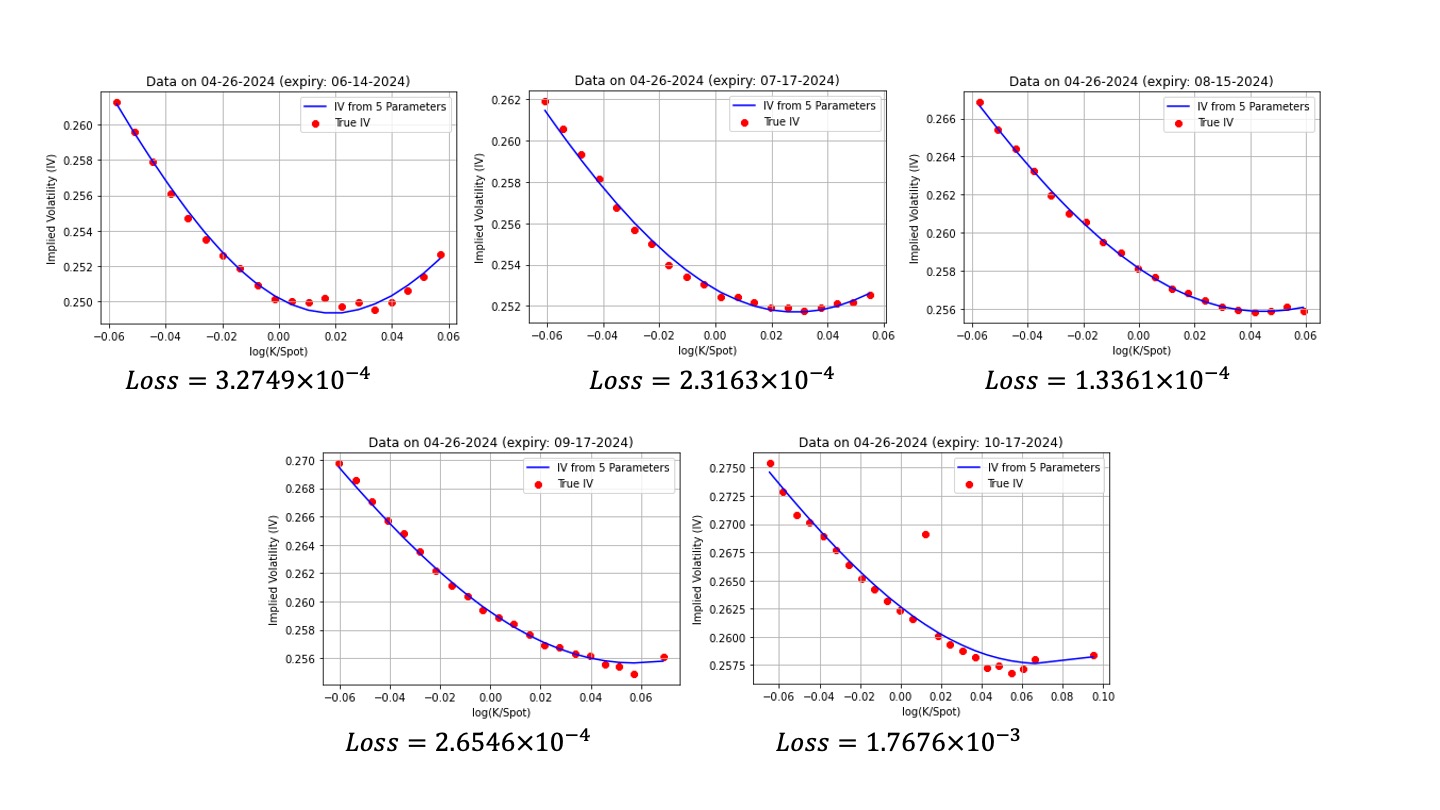}
\caption{True and calibrated IV for contracts of 5 expiry dates with loss values of data on 04-26-2024, fixing $\bar{v}=0.0763$ and $\lambda=0.45$.}
\label{fig:calib_0426_fix2}
\end{figure}

Since the effect of $\bar{v}$ and $\lambda$ on IV curves is much smaller than and can be offset by $v_0$ and $\eta$, we now fix $\bar{v}$ and $\lambda$ constant and regard the remaining 3 parameters as variables. We fix $\lambda=0.45$, which is the overall level of $\lambda$ in Table \ref{tab:calib_fix0}, and $\bar{v}=0.0763$, which is the average of all $\bar{v}$ in Table \ref{tab:calib_fix0}. Figure \ref{fig:calib_0424_fix2} to \ref{fig:calib_0426_fix2} display the true and fitted IVs of the 5 option contracts on 3 days, and Table \ref{tab:calib_fix2} shows the calibrated 5 parameters of the 5 contracts on 3 days when fixing $\bar{v}=0.0763$ and $\lambda=0.45$.

Table \ref{tab:calib_fix2} below summaries the corresponding $v_0$, $\eta$, and $\rho$ learned for each contract on the selected dates, based on fixed $\bar{v}$ and $\lambda$.

\begin{table}[H]
    \centering
    \begin{tabular}{|l||*{5}{c|}}
    \hline
    \diagbox[width=6em]{Date}{Expiry} &\makebox[3.5em]{06-14}  &\makebox[3.5em]{07-17} & \makebox[3.5em]{08-15} & \makebox[3.5em]{09-17} & \makebox[3.5em]{10-17} \\\hline\hline
    04-24 & \begin{tabular}{@{}c@{}c@{}}$v_0=0.0756$ \\ $\eta=0.9072$ \\ $\rho=-0.1614$\end{tabular} & \begin{tabular}{@{}c@{}c@{}}$v_0=0.0753$ \\ $\eta=0.7017$ \\ $\rho=-0.1536$\end{tabular} & \begin{tabular}{@{}c@{}c@{}}$v_0=0.0784$ \\ $\eta=0.6314$ \\ $\rho=-0.1799$\end{tabular} & \begin{tabular}{@{}c@{}c@{}}$v_0=0.0794$ \\ $\eta=0.6026$ \\ $\rho=-0.2045$\end{tabular} & \begin{tabular}{@{}c@{}c@{}}$v_0=0.0843$ \\ $\eta=0.6535$ \\ $\rho=-0.2067$\end{tabular} \\\hline\hline
    04-25 & \begin{tabular}{@{}c@{}c@{}}$v_0=0.0723$ \\ $\eta=0.8329$ \\ $\rho=0.0314$\end{tabular} & \begin{tabular}{@{}c@{}c@{}}$v_0=0.0721$ \\ $\eta=0.6601$ \\ $\rho=-0.0147$\end{tabular} & \begin{tabular}{@{}c@{}c@{}}$v_0=0.0751$ \\ $\eta=0.5998$ \\ $\rho=-0.0720$\end{tabular} & \begin{tabular}{@{}c@{}c@{}}$v_0=0.0783$ \\ $\eta=0.6417$ \\ $\rho=-0.1212$\end{tabular} & \begin{tabular}{@{}c@{}c@{}}$v_0=0.0816$ \\ $\eta=0.6413$ \\ $\rho=-0.1536$\end{tabular} \\\hline\hline
    04-26 & \begin{tabular}{@{}c@{}c@{}}$v_0=0.0703$ \\ $\eta=0.8845$ \\ $\rho=-0.0729$\end{tabular} & \begin{tabular}{@{}c@{}c@{}}$v_0=0.0716$ \\ $\eta=0.7003$ \\ $\rho=-0.0801$\end{tabular} & \begin{tabular}{@{}c@{}c@{}}$v_0=0.0760$ \\ $\eta=0.6696$ \\ $\rho=-0.1199$\end{tabular} & \begin{tabular}{@{}c@{}c@{}}$v_0=0.0793$ \\ $\eta=0.6807$ \\ $\rho=-0.1523$\end{tabular} & \begin{tabular}{@{}c@{}c@{}}$v_0=0.0841$ \\ $\eta=0.6980$ \\ $\rho=-0.1809$\end{tabular} \\\hline
    \end{tabular}
    \caption{Calibrated parameters of contracts of 5 expiry dates on 3 days (all dates in year 2024), fixing $\bar{v}=0.0763$ and $\lambda=0.45$.}
    \label{tab:calib_fix2}
\end{table}

The first observation is that $v_0$ in Table \ref{tab:calib_fix2} becomes more unstable than those in Table \ref{tab:calib_fix0}. This is because after fixing $\bar{v}$ as a constant, $v_0$ need to be more diverged in order to offset the difference caused by fixing $\bar{v}$. Hence, although $v_0$ is relatively stable among the same option contract on different days, it is quite unstable among option contracts with different time to maturities.

The second observation is that the term structures of $\eta$ and $\rho$ in Table \ref{tab:calib_fix2} demonstrate similar trends as from Table \ref{tab:calib_fix0}. Such pattern supports the validness of Heston model under the WTI calibration.

\newpage
\section{Conclusion}
\label{sec:conclusion}

In conclusion, we derive the Heston Model for option pricing, analyzed the extreme cases, and verify the results with simulation. We study the Greeks based on both analytical and numerical results. Using Crude Monte Carlo and Mixing Monte Carlo simulations, we test the Greek letters and plot the implied volatility curves for $\lambda$, $\eta$, and $\rho$.

We additionally evaluate the Heston model with real market data for WTI crude future and option contracts. Machine learning models such as Gradient Descent are utilized to help optimize the loss function and learn the optimal parameters on select dates. Our results show that neither $v_0$, $\eta$, or $\rho$ remains constant among options with different time to maturities or on different days. Hence, assuming constant parameters like $\eta$ and $\rho$ is disagreed by empirical market data and hence is a limitation of Heston model.

In future studies, we aim to develop better machine learning models. Instead of MSE to establish the loss, we will enforce rewards for correct directional predictions of the price and volatility for real trading purposes.

The current project has identified several limitations that suggest areas for improvement and further exploration. Future work could address these issues to enhance the robustness and applicability of the model:

\begin{enumerate}
    \item \textbf{Data Quality Improvement:} Some records from Barchart display an implied volatility (IV) of $0.00\%$, which likely indicates a system error. Future studies should include mechanisms to validate and correct such data anomalies.
    
    \item \textbf{Data Expansion:} Currently, the dataset is limited to only six data points obtained from Barchart. Expanding the dataset will improve the statistical significance of the model and allow for more comprehensive validation and testing.

    \item \textbf{Commodity Selection:} Crude Oil WTI behaves a highly volatile with respect to the recent war crisis. The unstable nature may negatively affect the parameterization by causing outliers.
    
    \item \textbf{Increase Iteration Count:} The algorithm is currently limited to a maximum of 300 iterations. Increasing the number of iterations could lead to more refined and accurate model parameters.
    
    \item \textbf{Alternative Loss Functions:} There is a need to explore alternative loss functions that prioritize predictive accuracy over the minimization of squared errors. Implementing a bonus rewarding system for correct directional predictions could foster models that are more aligned with practical trading scenarios.
\end{enumerate}

For the future, it is worth investigating that how to adjust the Heston model and modify the algorithm to produce optimal constant parameters for each contract. Crude Oil WTI's chaotic nature may not be a good suit to fully represent the essences of the Heston model for option pricing.

Profiting from exploring the arbitrage opportunites based on the outlier data through the IV smile curves can potentially be a valid research direction and help answer why the Crude Oil WTI fails the Heston cosntant parameters.

\newpage
\section{Acknowledgement}

We express our deepest gratitude to Professor Roza Galeeva for her invaluable teaching and mentorship throughout the duration of this project. Her guidance over the past year has been instrumental in shaping our understanding and approach, providing us with the tools and insights needed to succeed. We are immensely thankful for her dedication and the significant impact she has had on our academic journey.

\newpage
\printbibliography

\newpage 
\section*{Appendix}
\subsection*{Appendix A: Characteristic Function}

Based on Heston's work in 1993\cite{heston1993closed}, this section introduces the methodology of Characteristic Function in the realm of solving for the Heston Model's solution. 

\dots

By analogy with the Black-Scholes formula, we guess a solution of the form
\begin{equation}\label{eqn:A1}
C(S, v, t) = SP_1 - KP(t, T)P_2, \tag{A1}
\end{equation}
where the first term is the present value of the spot asset upon optimal exercise, and the second term is the present value of the strike-price payment. Both of these terms must satisfy the original PDE \eqref{eqn:hst_pde}. It is convenient to write them in terms of the logarithm of the spot price
\begin{equation}
x = \ln[S]. \tag{A2}
\end{equation}

Substituting the proposed solution \eqref{eqn:A1} into the original PDE \eqref{eqn:hst_pde} shows that \( P_1 \) and \( P_2 \) must satisfy the PDEs
\begin{equation}\label{eqn:A3}
\frac{1}{2} v^2 \frac{\partial^2 P_j}{\partial x^2} + \rho v \frac{\partial^2 P_j}{\partial x \partial v} + \frac{1}{2} \sigma^2 v^2 \frac{\partial^2 P_j}{\partial v^2} + (r + \mu v) \frac{\partial P_j}{\partial x} + (a - b_j v) \frac{\partial P_j}{\partial v} + \frac{\partial P_j}{\partial t} = 0, \tag{A3}
\end{equation}
for \( j = 1,2 \), where
\( u_1 = \frac{1}{2}, u_2 = -\frac{1}{2}, a = k\theta, b_1 = k + \lambda - \rho \sigma, b_2 = k + \lambda \).

For the option price to satisfy the terminal condition in Equation \eqref{eqn:terminal_P}, these PDEs \eqref{eqn:A3} are subject to the terminal condition

\begin{equation}
    P_j(x, v, T; \ln[K]) = 1_{x \geq ln(k)}\tag{A4}
\end{equation}

we may also write this indicator function into its probability formula:

\begin{equation}
P_j(x, v, T; \ln[K]) = \frac{1}{2} \pm \frac{1}{2} \mathrm{erf}\left(\frac{x-\ln[K]}{\sqrt{2T}}\right). \tag{A5}
\end{equation}

% check if above is correct

Thus, they may be interpreted as ``adjusted" or ``risk-neutralized" probabilities (See \cite{cox1976valuation}). The Appendix explains that when \( x \) follows the stochastic process

\begin{equation}\begin{aligned}
dx(t) &= (r + u_j v)dt + \sqrt{v(t)} dz_1(t),\\
dv &= (a_j - b_j v)dt + \sigma \sqrt{v(t)} dz_2(t), 
\end{aligned}\tag{A6}\end{equation}
where the parameters \( u_j, a_j, \) and \( b_j \) are defined as before, then \( P_j \) is the conditional probability that the option expires in-the-money:
\begin{equation}
P_j(x, v, T; \ln[K]) = \Pr[x(T) \geq \ln[K] \mid x(t) = x, v(t) = v]. \tag{A7}
\end{equation}
The probabilities are not immediately available in closed form. However, the Appendix shows that their characteristic functions, \( f_1(x, v, T; \phi) \) and \( f_2(x, v, T; \phi) \) respectively, satisfy the same PDEs \eqref{eqn:A3}, subject to the terminal condition
\begin{equation}
f_j(x, v, T; \phi) = e^{i\phi x}. \tag{A8}
\end{equation}

\begin{equation}\label{eqn:A9}
f(x, v, t; \phi) = e^{C(T-t;\phi) + D(T-t;\phi)v + \phi x}, \tag{A9}
\end{equation}

where

\begin{align*}
C(T; \phi) &= \phi rT + \frac{a}{\sigma^2} \left[ (b_j - \rho \sigma \phi + d)T - 2 \ln \left( \frac{1 - ge^{dT}}{1 - g} \right) \right], \\
D(T; \phi) &= \frac{b_j - \rho \sigma \phi + d}{\sigma^2} \left[ 1 - \frac{1 - e^{dT}}{1 - ge^{dT}} \right], \\
\end{align*}

and

\begin{align*}
g &= \frac{b_j - \rho \sigma \phi + d}{b_j - \rho \sigma \phi - d'}, \\
d &= \sqrt{(\rho \sigma \phi - b_j)^2 - \sigma^2(2 u_j \phi - \phi^2)}.
\end{align*}

One can invert the characteristic functions to get the desired probabilities:

\begin{equation}\label{eqn:A10}
P(x, v, T; ln[K]) = \frac{1}{2} + \frac{1}{\pi} \int_0^\infty \text{Re} \left\{ \frac{e^{-i\phi ln[K]}f(x, v, T; \phi)}{i\phi} \right\} d\phi. \tag{A10}
\end{equation}

The integrand in Equation (18) is a smooth function that decays rapidly and presents no difficulties.

Equations \eqref{eqn:A1}, \eqref{eqn:A9}, and \eqref{eqn:A10} give the solution for European call options. In general, one cannot eliminate the integrals in Equation \eqref{eqn:A10}, even in the Black-Scholes case. However, they can be evaluated in a fraction of a second on a microcomputer by using approximations similar to the standard ones used to evaluate cumulative normal probabilities.

\subsection*{Appendix B: Heston Model for American Option}

Based on American option pricing under stochastic volatility: an efficient numerical approach by Farid AitSahlia, Manisha Goswami, and Suchandan Guha, 
\url{https://bear.warrington.ufl.edu/aitsahlia/AitSahlia_CMS_1.pdf}

For American call option we will make the following modification for the pricing model, as compared to the European call option. Considering strike \( K \). Let \( C_A(S, v, \tau) \) denote its price when the underlying has price \( S \) and spot volatility \( v \), with \( \tau \) units of time left to expiry. Using standard arbitrage arguments, \( C_A \) can be shown to satisfy the following partial differential equation
\begin{equation}\label{eqn:B1}
\frac{\partial C_A}{\partial \tau} = \frac{1}{2} v^2 S^2 \frac{\partial^2 C_A}{\partial S^2} + \rho v S \frac{\partial^2 C_A}{\partial S \partial v} + \frac{1}{2} \sigma^2 v^2 \frac{\partial^2 C_A}{\partial v^2} + (r - q)S \frac{\partial C_A}{\partial S} + (\kappa(\theta - v) - \lambda v) \frac{\partial C_A}{\partial v} - rC_A \tag{B1}
\end{equation}
in the region \( D = \{0 \leq \tau \leq T, 0 \leq S \leq b(v, \tau), 0 < v < \infty\} \) along with the boundary conditions
\begin{align*}
C_A(S, v, 0) &= \max(S - K, 0), \\
C_A(b(v, \tau), v, \tau) &= b(v, \tau) - K, \\
\lim_{S \to b(v, \tau)} \frac{\partial C_A}{\partial S} &= 1, \\
\lim_{S \to b(v, \tau)} \frac{\partial C_A}{\partial v} &= 0,
\end{align*}
where \( b(v, \tau) \) denotes the optimal early exercise price (boundary) at time \( \tau \) for spot volatility \( v \), and \( \lambda v \) denotes the corresponding market price of volatility risk, with \( \lambda \) determined empirically. 

Recall, as suggested by Heston in 1993, the approach to calculate the market price of risk is designed to address the incompleteness of the market information inherented in the stochastic volatility modeling.

Chiarella and Ziogas \cite{chiarella2005pricing} use the method of Jamshidian \cite{jamshidian1992analysis} to convert the
homogeneous PDE \eqref{eqn:B1} defined in the region D above to an inhomogeneous one in an unrestricted domain.

For illustrative purposes, we consider an American call option with strike \( K \). Let \( C_A(S, v, \tau) \) denote its price when the underlying has price \( S \) and spot volatility \( v \), with \( \tau \) units of time left to expiry. Using standard arbitrage arguments, \( C_A \) can be shown to satisfy the following partial differential equation

\begin{equation}\label{eqn:B2}\begin{aligned}
    \frac{\partial C_A}{\partial \tau} =& \frac{v^2}{2} \frac{\partial^2 C_A}{\partial x^2} + \rho v \frac{\partial^2 C_A}{\partial x \partial v} + \frac{\sigma^2 v}{2} \frac{\partial^2 C_A}{\partial v^2} + \left(r - q - \frac{v}{2}\right) \frac{\partial C_A}{\partial x} \\
    &+ (\alpha - \beta v) \frac{\partial C_A}{\partial v} - H(x - \ln b(v, \tau))e^{r\tau}(qe^{qx} - rK), 
\end{aligned}\tag{B2}\end{equation}

where \( \alpha = k\theta \) and \( \beta = k + \lambda \), in the unrestricted domain \( -\infty < x < \infty \), \( 0 < v < \infty \), \( 0 \leq \tau \leq T \), subject to the boundary conditions:
\begin{align*}
C_A(x, v, 0) &= \max(e^x - K, 0), \\
\lim_{x \to -\ln b(v, \tau)} \frac{\partial C_A}{\partial x} &= b(v, \tau)e^{r\tau}, \\
\lim_{x \to -\ln b(v, \tau)} \frac{\partial C_A}{\partial v} &= 0,
\end{align*}
where \( H(x) \) is the Heaviside step function defined as
\[
H(x) = 
\begin{cases}
1, & x > 0, \\
\frac{1}{2}, & x = 0, \\
0, & x < 0.
\end{cases}
\]

To obtain \( C_A \) through Eq. \eqref{eqn:B2}, one still needs the knowledge of the optimal stopping (early exercise) boundary \( b(v, \tau) \). In the classical context of constant volatility for the underlying asset return, A\"it-Sahalia and Lai \cite{aitsahlia2000canonical} have shown that this boundary is well-approximated by linear splines with very few knots, typically 3 or 4. When the volatility of the underlying asset itself follows a stochastic process as in Equation \eqref{eqn:B1} above, Broadie et al. \cite{broadie2000pricing} produced empirical evidence to suggest that the corresponding optimal stopping surface can be well-approximated in a log-linear fashion near the long-term variance level; i.e.:
\[
\ln b(v, \tau) \approx b_0(\tau) + vb_1(\tau), \quad \text{for } v \text{ near } \theta,
\]
thus reducing the determination of \( b(v, \tau) \) to that of \( b_0(\tau) \) and \( b_1(\tau) \). Under this assumption, Chiarella and Ziogas (2005) then express the solution for the PDE \eqref{eqn:B2} as the following decomposition formula:

\begin{equation}\begin{aligned}
C_A(S, v, \tau) =& S e^{-q\tau} P_1(S, v, \tau, K; 0) - Ke^{-r\tau} P_2(S, v, \tau, K; 0) \\
&+ \int_0^\tau S e^{-q(\tau-\xi)} P_1(S, v, \tau - \xi, e^{b_0(\xi)}, -b_1(\xi))d\xi \\
&- \int_0^\tau K e^{-r(\tau-\xi)} P_2(S, v, \tau - \xi, e^{b_0(\xi)}, -b_1(\xi))d\xi, 
\end{aligned}\tag{B3}\end{equation}

where
\begin{equation}
P_j(\mathcal{S}, v, \tau - \xi; b; w) = \frac{1}{2} + \frac{1}{\pi} \int_0^{\infty} \text{Re} \left( \frac{f_j(\mathcal{S}, v, T - \xi; \phi, w)e^{-i\phi \ln b}}{i\phi} \right) d\phi, \tag{B4}
\end{equation}
for \( j = 1, 2 \) and
\begin{align*}
f_1(\mathcal{S}, v, \tau - \xi; \phi, w) &= e^{-\ln S}e^{-(r-q)(\tau-\xi)} f_2(\mathcal{S}, v, T - \xi; \phi, w), \\
f_2(x, v, \tau - \xi; \phi, \psi) &= \exp\left[g_0(\phi, \psi, \tau - \xi) + g_1(\phi, \psi, \tau - \xi)x + g_2(\phi, \psi, \tau - \xi)v\right],
\end{align*}
with
\begin{align*}
g_0(\phi, \psi, \tau - \xi) &= (r - q)i\phi(\tau - \xi) + \frac{\alpha}{\sigma^2} \left[(\beta - \rho i\phi + D_2)(\tau - \xi) - 2 \ln \left( \frac{1 - G_2(\psi)e^{D_2(\tau-\xi)}}{1 - G_2(\psi)} \right)\right], \\
g_1(\phi, \psi, \tau - \xi) &= i\phi, \\
g_2(\phi, \psi, \tau - \xi) &= i\psi + \frac{\beta - \rho i\phi - \sigma^2 i\psi + D_2}{\sigma^2} \left[ 1 - \frac{1 - e^{D_2(\tau-\xi)}}{1 - G_2(\psi)e^{D_2(\tau-\xi)}} \right],
\end{align*}
where \( D_2 \) is defined as
\begin{equation}
D_2^2 \equiv (\rho i\phi - \beta)^2 + \sigma^2(\phi + i)^2,\tag{B5}
\end{equation}
and
\begin{equation}
G_2(\psi) \equiv \frac{\beta - \rho i\phi - \sigma^2 i\psi + D_2}{\beta - \rho i\phi - \sigma^2 i\psi - D_2}.\tag{B6}
\end{equation}

\subsection*{Appendix C: Dynamics of Black-Scholes Model}

The Black-Scholes (BS) model, also known as the Black-Scholes-Merton model, is a mathematical framework for pricing European-style options and similar financial instruments \cite{black1973pricing}. Unlike the Heston model, which incorporates stochastic volatility, the Black-Scholes model assumes that the volatility of the underlying asset is constant and the returns of the asset are normally distributed. The dynamics of the underlying asset's price \( S_t \) in the Black-Scholes model are given by the following stochastic differential equation (SDE):

\[
dS_t = \mu S_t \, dt + \sigma S_t \, dW_t
\]

where:
- \( S_t \) is the stock price at time \( t \).
- \( \mu \) is the expected return (drift) of the stock.
- \( \sigma \) is the constant volatility of the stock's returns.
- \( W_t \) is a standard Brownian motion (or Wiener process).

This model underpins much of the theory used in the pricing of derivatives and has been a foundational element in financial economics.

\subsection*{Appendix D: Alternative Ways of Calculating Greeks}

The discounted payoff of a European call option with strike \( K \) and maturity \( T \) is given by \( e^{-rT}(S_T - K)^+ \), where \( S_T \) is the stock price at time \( T \). In the equations below \( \mathbf{1}_A \) is used to denote the indicator function of the event \( A \). PW derivative estimators for a European call option are given in \eqref{eqn:D1} and \eqref{eqn:D2} below.

PW estimators:
\begin{equation}\label{eqn:D1}
\text{Delta: } e^{-rT} \mathbf{1}_{\{S_T \geq K\}}\frac{S_T}{S_0} \tag{D1}
\end{equation}
\begin{equation}\label{eqn:D2}
\text{Rho: } e^{-rT} \mathbf{1}_{\{S_T \geq K\}}KT \tag{D2}
\end{equation}

To derive the LR estimators, we need the conditional density of \( S_T \).This density can be written as:
\[
g(x) = \frac{1}{x\sigma\sqrt{T}} \phi(d(x)),
\]
where \( \phi(\cdot) \) is the standard normal density function and
\[
d(x) = \frac{\ln(x/(S_0e^{(r - \frac{1}{2}\sigma^2)T}))}{\sigma\sqrt{T}}.
\]
To find the delta estimator, we first take the derivative with respect to \( S_0 \). After some simplification, we get:
\[
\frac{\partial g(x)}{\partial S_0} = -\frac{d(x)\phi(d(x))}{xS_0\sigma^2T}.
\]
Dividing this by \( g(x) \) and evaluating the expression at \( x = S_T \) gives the score function for LR delta estimator:
\[
\frac{\partial g(S_T)/\partial S_0}{g(S_T)} = -\frac{d(S_T)}{S_0\sigma\sqrt{T}}.
\]

Other estimators can be derived in a similar fashion. For details, see Broadie and Glasserman \cite{broadie1996estimating}.

LR estimators:
\begin{equation}\label{eqn:D3}
\text{Delta: } e^{-rT}(S_T - K)^+\left(\frac{d}{S_0\sigma\sqrt{T}}\right) \tag{D3}
\end{equation}
\begin{equation}\label{eqn:D4}
\text{Gamma: } e^{-rT}(S_T - K)^+\left(\frac{d^2 - d\sigma\sqrt{T} - 1}{S_0^2\sigma^2T}\right) \tag{D4}
\end{equation}
\begin{equation}\label{eqn:D5}
\text{Rho: } e^{-rT}(S_T - K)^+\left(-T + \frac{d\sqrt{T}}{\sigma}\right) \tag{D5}
\end{equation}
where \( d = \left(\ln(S_T/(S_0e^{(r - \frac{1}{2}\sigma^2)T}))\right)/(\sigma\sqrt{T}) \) in \eqref{eqn:D3}--\eqref{eqn:D5}. If \( S_T \) is generated from \( S_0 \) using a normal random variable \( Z \), then \( d = Z \), and these estimators are easily computed in a simulation.

The delta estimator in \eqref{eqn:D3} includes an indicator function, so the PW method cannot be used to take the derivative
of this expression to obtain a gamma estimator. For finding
estimators for second order derivatives like gamma, we can
use a mixed estimator where we use the PW method for
one order of differentiation and LR method for the other.

This gives the estimators in \eqref{eqn:D6} and \eqref{eqn:D7} for the gamma
of a European call option.

Mixed estimators:
\begin{equation}\label{eqn:D6}
\text{LR-PW Gamma: } e^{-rT} \mathbf{1}_{\{S_T \geq K\}}K \left( \frac{d}{S_0^2\sigma\sqrt{T}} \right) \tag{D6}
\end{equation}
\begin{equation}\label{eqn:D7}
\text{PW-LR Gamma: } e^{-rT} \mathbf{1}_{\{S_T \geq K\}}\frac{S_T}{S_0^2} \left( \frac{d}{\sigma\sqrt{T}} - 1 \right) \tag{D7}
\end{equation}
where \( d \) is as given above.

If we assume that the correlation \( \rho \) is constant, we can calculate the sensitivity to it, rhoza. We get
\begin{equation}\label{eqn:D8}
\frac{\partial C}{\partial \rho} = -\frac{\rho}{1 - \rho^2} \left( \frac{\partial C}{\partial \sigma_{\text{eff}}} + F_{\text{eff}} \frac{\partial C}{\partial F_{\text{eff}}} \frac{\partial Y(T)}{\partial \rho} \right) \tag{D8}
\end{equation}
where
\begin{equation}\label{eqn:D9}
\frac{\partial Y(T)}{\partial \rho} = -\rho \int_{0}^{T} \sigma^2(t)dt + \int_{0}^{T} \sigma(t)dW_2(t) \tag{D9}
\end{equation}

Thus, the total sensitivity to correlation \( \text{Rh} \) comes from the changes in the effective volatility \( \sigma_{\text{eff}} \), the Vega part \( \text{Rh}^\sigma \), and the changes in the effective price \( F_{\text{eff}} \), the Delta part, \( \text{Rh}^F \). \( \text{Rh}^\sigma \) and \( \text{Rh}^F \) are of the opposite signs, depending on the correlation sign and the moneyness. In Appendix A, we exemplify these contributions using numerical tests for the Heston model.

The mixing theorem and all derived results are true not only for call/put options, but for any European style claims with arbitrary payoff function at expiry \( g(F(T)) \). The call option values \( c \) have to be replaced by the fair value of the claim \( f(F, V, T) \), see \cite{lewis2009option},
\begin{equation}\label{eqn:D10}
f(F, V, T) = e^{-rT} \left( F \exp\left(-\frac{1}{2} V_T \right) \sqrt{V_T} \right) \tag{D10}
\end{equation}
where the function \( J(a, b) \) is given by
\begin{equation}\label{eqn:D11}
J(a, b) = \frac{1}{\sqrt{2\pi}} \int_{-\infty}^{\infty} g(ae^{bx})\exp\left(-\frac{1}{2} x^2\right) dx \tag{D11}
\end{equation}

Roza Galeeva also provides \cite{rhoza2022}, 2022,

the numerical way to calculate rhoza for SV models, in particular, the Heston model,
\[
b(V(t)) = \lambda(\theta - V(t)), \quad a(V(t)) = \eta \sqrt{V(t)}
\]
where \( \lambda \) is the speed of reversion of \( V(t) \) to its long-term mean \( \theta \).

\subsection*{Appendix E: Algorithm for Gradient Descent Example}

\begin{algorithm}
\caption{Gradient Descent for Parameter Optimization}
\begin{algorithmic}[1]
\State \textbf{Input:} data frame $df$, data arrays $data\_arrays$, initial parameters $initial\_params$
\State \textbf{Parameters:} learning rates $initial\_learning\_rate$, decay level $learning\_rate\_decay\_lv$, denominator $learn\_deno$, iteration count $iterations$
\State $params \gets \text{np.array}(initial\_params, \text{dtype=np.float32})$
\State $epsilon \gets 0.0001$  \Comment{Small value for finite difference method}

\For{$i \gets 0$ \textbf{to} $iterations-1$}
    \State Initialize an empty list $gradients$
    \For{$j \gets 0$ \textbf{to} $\text{len}(params)-1$}
        \State $params\_up \gets params.\text{copy}()$
        \State $params\_up[j] \gets params\_up[j] + epsilon$
        \State $loss\_up \gets \text{loss\_function\_np\_new}(df, data\_arrays, params\_up)$
        
        \State $params\_down \gets params.\text{copy}()$
        \State $params\_down[j] \gets params\_down[j] - epsilon$
        \State $loss\_down \gets \text{loss\_function\_np\_new}(df, data\_arrays, params\_down)$
        
        \State $gradient \gets (loss\_up - loss\_down) / (2 \times epsilon)$
        \State Append $gradient$ to $gradients$
    \EndFor

    \State $learning\_rate \gets initial\_learning\_rate / (learn\_deno + i)^{learning\_rate\_decay\_lv}$
    \State $params \gets params - learning\_rate \times \text{np.array}(gradients)$
    \State Adjust $params$ within specific bounds
    \State $loss \gets \text{loss\_function\_np\_new}(df, data\_arrays, params)$
    \State Print the iteration, parameters, and loss
\EndFor

\State \textbf{return} $params$
\end{algorithmic}
\end{algorithm}

\end{document}